\documentstyle[12pt,a4,psfig]{article}
\def\xa{x_\alpha^{}}
\def\contact{g_E^2/c^2_\chi m_{Z_2}^2}
\def\xibar{\bar{\xi}}
\def\to{\rightarrow}

\def\ie{{\it i.e.}}
\def\eg{{\it e.g.}}

\def\vsp#1{\vspace{#1 cm}}
\def\hsp#1{\hspace{#1 cm}}
\def\ov{\overline}
\def\disp{\displaystyle}
\def\gsim{~{\rlap{\lower 3.5pt\hbox{$\mathchar\sim$}}\raise 1pt\hbox{$>$}}\,}
\def\lsim{~{\rlap{\lower 3.5pt\hbox{$\mathchar\sim$}}\raise 1pt\hbox{$<$}}\,}

\def\gm5{\gamma_5}

\def\mmz{m_Z^2}
\def\hph{\hphantom{-}}
\def\lenc{L.E.N.C.\ }
\def\mt{m_t^{}}
\def\mh{m_H^{}}
\def\xs{x_s^{}}
\def\xt{x_t^{}}
\def\xh{x_H^{}}
\def\mh{m_H^{}}

\def\ds{\Delta S}
\def\dt{\Delta T}
\def\du{\Delta U}

\def\sbar{\bar{s}^2}

\def\gzbar{\bar{g}_Z^2}
\def\gwbar{\bar{g}_W^2}
\def\abar{\bar{\alpha}}

\def\dgzbar{\Delta \gzbar}
\def\dsbar{\Delta \sbar}

\def\etal{{\it et al.~}}
\newcommand{\beq}{\begin{equation}}
\newcommand{\eeq}{\end{equation}}
\newcommand{\bea}{\begin{eqnarray}}
\newcommand{\eea}{\end{eqnarray}}
\newcommand{\bsub}{\begin{subequations}}
\newcommand{\esub}{\end{subequations}}

\renewcommand{\theequation}{\thesection.\arabic{equation}}
\newcommand{\clean}{\setcounter{equation}{0}}

\def\esix_contents{
\begin{table}[t]
\begin{center}
\caption{
The hypercharge $Y$ and the ${\rm U(1)'}$ charge $Q_E$ of 
all the matter fields in a {\bf 27} for the $\chi, \psi, \eta$ 
and $\nu$ models. 
The classification of the fields in the SO(10) and the SU(5) 
groups is also shown. 
The value of ${\rm U(1)'}$ charge follows the hypercharge 
normalization. }
\begin{tabular}{cccccccc}  \hline \hline \\ [-0.4cm]
\vsp{0.2}
SO(10)& SU(5) & field & $Y$ & $2\sqrt{6}Q_\chi$ 
	& $\sqrt{72/5}Q_\psi$ & $Q_\eta$ 
	& $Q_\nu$ \\ \hline \\
\vsp{0.2}
{\bf 16} & {\bf 10} & 
$Q$     & $+\frac{1}{6}$ & $-1$& $+1$ & $-\frac{1}{3}$
	& $+\sqrt{\frac{1}{24}}$\\
& & 
$u^c$   & $-\frac{2}{3}$ & $-1$& $+1$& $-\frac{1}{3}$ 
	& $+\sqrt{\frac{1}{24}}$ \\
& & 
$e^c$   & $+1 $ & $-1$& $+1$& $-\frac{1}{3}$
	& $+\sqrt{\frac{1}{24}}$\\
\\
& $\bf{\ov{5}}$ &
$L$     & $-\frac{1}{2}$& +3 & $+1$& $+\frac{1}{6}$
	& $+\sqrt{\frac{1}{6}}$\\
& &
$d^c$   & $+\frac{1}{3}$ & $+3$& $+1$& $+\frac{1}{6}$
	& $+\sqrt{\frac{1}{6}}$ \\ 
\\
& {\bf 1} &
$\nu^c$ & $0$ & $-5$ & $+1$& $-\frac{5}{6}$& 0 \\ 
\\
{\bf 10} &  $\bf{5}$ &
$H_u$   & $+\frac{1}{2}$& +2 & $-2$& $+\frac{2}{3}$
	& $-\sqrt{\frac{1}{6}}$\\
& &
$D$     & $-\frac{1}{3}$ & $+2$& $-2$& $+\frac{2}{3}$
	& $-\sqrt{\frac{1}{6}}$ \\ 
\\ 
& $\bf{\ov{5}}$ &
$H_d$   & $-\frac{1}{2}$ & $-2$ & $-2$& $+\frac{1}{6}$
	& $-\sqrt{\frac{3}{8}}$\\
& &
$\ov{D}$& $+\frac{1}{3}$ & $-2$ & $-2$& $+\frac{1}{6}$
	& $-\sqrt{\frac{3}{8}}$ \\ 
\\
{\bf 1} & {\bf 1} & 
$S$     & $0$ & $0$ & $4$ & $-\frac{5}{6}$
	& $\sqrt{\frac{25}{24}}$ \\ 
\hline 
\hline 
\end{tabular} 
\end{center}
\label{table_u1charge}
\end{table}
}
\def\lep_table{
	\begin{table}[tp]
	\begin{center}
	\begin{tabular}{|r|c|r|r|r|r|r|r|}
	\hline
	 \multicolumn{2}{|c}{}& \multicolumn{6}{|c|}
	{pull = $\frac{\langle {\rm data} \rangle - {\rm best~fit}}
	{\langle {\rm error} \rangle}$} \\ 
\cline{3-8}
	 \multicolumn{2}{|c|}{}& SM & $\chi$ & $\psi$ & $\eta$ 
	& $\nu$ & $\eta^*$ 	\\ \hline
	\multicolumn{2}{|c|}{$Z$-pole experiments}&&&&&& \\ \hline
	$m^{}_Z$ (GeV) & 91.1867$\pm$0.0020&&&&&&\\ 
	$\Gamma_Z^{}$ (GeV)  & 2.4948  $\pm$ 0.0025  
	& $-0.8$ & $-0.8$ & $-0.6$ & $-0.7$ &$-0.7$& $-0.6$\\
	$\sigma^0_h$(nb) & 41.486 $\pm$ 0.053 
	& 0.3 & 0.6 & 0.6 & 0.3 & 0.6 & 0.2\\ 
	$R_{\ell}$ & 20.775 $\pm $ 0.027 
	& 0.9 & 0.8 & 0.7 & 0.9 & 0.7 & 1.1\\
	$A^{0,\ell}_{FB}$ & 0.0171 $\pm $ 0.0010 
	& 0.8 & 0.8 & 0.7 & 0.7 & 0.8& 0.7 \\
	$A_{\tau}$& 0.1411 $\pm$ 0.0064 
	& $-$1.0 & $-1.0$ & $-1.0$ & $-1.0$ & $-1.0$& $-1.0$ \\
	$A_{e} $  & 0.1399 $\pm$ 0.0073 
	& $-$1.1 & $-1.0$ & $-1.1$ & $-1.1$ &$-1.1$& $-1.1$ \\
	$R_b$ & 0.2170 $\pm$ 0.0009 
	& 1.4 & 1.4 & 1.5 & 1.4 & 1.4 & 0.6 \\
	$R_c$ & 0.1734 $\pm$ 0.0048 
	& 0.3 & 0.3 & 0.2 & 0.3 & 0.3 & 0.5 \\
	$A^{0,b}_{FB}$ & 0.0984 $\pm$ 0.0024 
	& $-$2.1 & $-2.0$ & $-2.1$ & $-2.1$ &$-2.1$& $-2.2$\\
	$A^{0,c}_{FB}$ & 0.0741 $\pm$ 0.0048 
	& 0.0 & 0.1 & $0.0$ & $0.0$ & $0.0$& $-0.1$ \\
	$A^0_{LR}$ & 0.1547 $\pm$ 0.0032 
	& 2.2 & 2.3 & 2.2 & 2.2 &2.2 & 2.2 \\
	$A_b$ & 0.900 $\pm$ 0.050 
	& $-$0.7 & $-$0.7 & $-$0.7 & $-$0.7 &$-0.7$& $-0.7$ \\
	$A_c$ & 0.650 $\pm$ 0.058 
	& $-$0.3 & $-$0.3 & $-$0.3 & $-$0.3 & $-0.3$& $-0.4$\\ \hline
	\multicolumn{2}{|c|}{$W$-mass measurement} &&&&&& \\ \hline
	$m^{}_W$ (GeV)&80.43 $\pm$ 0.084
	&0.5 &0.5 &0.5 &0.5 & 0.5& 0.5
	\\ \hline \hline
	\multicolumn{2}{|c|}{$\chi_{\rm min}^2$ and d.o.f.}
	&&&&&& \\ \hline
	$\chi^2_{\rm min}$ & 
	& 16.9 & 16.7 & 16.7 & 16.9 & 16.6 & 16.1\\
	d.o.f. & &14  &12  &12  &12 &12 & 12\\ 
	\hline \hline
	parameters & constraints & 
	\multicolumn{6}{|c|}{best fit values } \\ \hline
	$m^{}_t~{\rm (GeV)}$ & $175.6 \pm 5.5$ 
	& 172.4 & 173.1 & 172.8 & 172.3 & 172.9 & 172.9\\
	$\alpha^{}_s(m_{Z_1})$ &$0.118\pm 0.003$ 
	& 0.1185 & 0.1179 & 0.1180 & 0.1185 & 0.1179 & 0.1192\\
	$1/\abar (m^2_{Z_1})$ & $128.75\pm 0.09$ 
	& 128.75 & 128.76 & 128.74 & 128.74 & 128.75 & 128.74 \\
	$T_{\rm new}$ &----- & ----- & 0 & 0 & 0 & 0 & 0\\
	$\bar{\xi}$   &----- & ----- & 0.0002 & 0.0002 & $-0.0001$ 
		& 0.0002 & 0.0027 \\
	\hline
	\end{tabular}
	\caption{\small 
	Summary of electroweak measurements for the $Z$-pole 
	experiments and the $m_W$ measurement~\cite{lep_slc_97}. 
	The best fits to all the data in this Table are found 
	by allowing the five parameters $\mt, \alpha_s(m_{Z_1}), 
	\abar(m_{Z_1}^2), T_{\rm new}$ and $\xibar$ to vary 
	freely under the constraints  
	$\mt = 175.6 \pm 5.5~{\rm GeV}$~~\cite{mt96}, 
	$\alpha_s(m_{Z_1}) = 0.118 \pm 0.003$~\cite{PDG}, 
	$1/\abar (m^2_{Z_1}) = 128.75\pm 0.09$~\cite{eidelman}, 
	$T_{\rm new} \geq 0$ and $\mh = 100~{\rm GeV}$. 
	The results for the $\chi, \psi, \eta$ and $\nu$ models are 
	obtained by setting $\delta = 0$. 
	The symbol $\eta^*$ denotes the leptophobic $\eta$-model 
	where $\delta$ is taken to be $\delta = 1/3$. 
	}
	\end{center}
	\label{table:lep97}
	\end{table}
	} 
\def\lenc_table{
	\begin{table}[tp]
	\begin{center}
	\begin{tabular}{|r|c|r|r|r|r|r|r|}
	\hline
	 \multicolumn{2}{|c}{}& \multicolumn{6}{|c|}
	{pull = $\frac{\langle {\rm data} \rangle - {\rm best\; fit}}
	{\langle {\rm error} \rangle}$} \\ 
\cline{3-8}
	 \multicolumn{2}{|c|}{}& SM & $\chi$ & $\psi$ & $\eta$ 
	& $\nu$ & $\eta^*$ 	\\ \hline
	\multicolumn{2}{|c|}{LENC experiments}&&&&&& \\ \hline
	$A_{\rm SLAC}$ & $0.80\;\;\;\; \pm 0.058\;$ 
		& 1.0 & 1.0 & 1.0 & 1.0 & 1.0 & 0.9 \\ 
	$A_{\rm CERN}$ & $-1.57 \;\;\;\,\pm 0.38\;\;\;\;\;\:$ 
		&$-0.4$ & $-0.4$ & $-0.4$ & $-0.4$ & $-0.4$ & $-0.4$ \\ 
	$A_{\rm Bates}$ & $-0.137\;\; \pm 0.033\;\;\;\;$ 
		& 0.5 & 0.4 & 0.4 & 0.4 & 0.4 & 0.5 \\ 
	$A_{\rm Mainz}$ & $-0.94\;\;\;\, \pm 0.19\;\;\;\;\;\;$ 
		& $-0.3$ & $-0.3$ & $-0.3$ & $-0.4$ & $-0.3$ & $-0.3$ \\ 
	$Q_W(^{133}_{55}C_s)$ & $-72.08\;\: \pm 0.92\;\;\;\;\;\;$ 
		& 1.0 & $-0.2$ & 1.0 & 0.2 & $-0.1$ & 1.3 \\ 
	$K_{FH}$  & $0.3247 \pm 0.0040$ 
		& $-1.5$ & $-1.4$ & $-1.5$ & $-1.5$ & $-1.4$ & $-1.4$ \\ 
	$K_{CCFR}$  & $0.5820 \pm 0.0049$ 
		& $-0.5$ & $-0.4$ & $-0.3$ & $-0.4$ & $-0.4$ & $-0.5$ \\ 
	$g_{LL}^{\nu_\mu e}$  & $-0.269\;\: \pm 0.011\;\;\;\;\:$ 
		& 0.4 & 0.1 & 0.1 & 0.5 & 0.1 & 0.4 \\ 
	$g_{LR}^{\nu_\mu e}$  & $\;\; 0.234\;\: \pm 0.011\;\;\;\:$ 
		& 0.1& 0.0 & 0.4 & 0.2 & 0.2 & 0.1\\ 
	\hline
	\multicolumn{2}{|c|}{$\chi^2_{\rm min}$ and d.o.f.}
	&&&&&& \\ \hline
	$\chi^2_{\rm min}$ & 
	& 22.0 & 20.2 & 21.5 & 21.2 & 20.4 & 21.7\\
	d.o.f. & & 23 & 20 & 20 & 20 & 20 & 21 \\ 
	\hline \hline
	parameters & constraints & 
	\multicolumn{6}{|c|}{best fit values} \\ \hline
	$m^{}_t~{\rm (GeV)}$ & $175.6 \pm 5.5$ 
	& 171.6 & 172.3 & 172.1 & 171.5 & 172.3 & 172.0 \\
	$\alpha_s(m_{Z_1})$ & $0.118 \pm 0.003$ 
	& 0.1185 & 0.1181 & 0.1181 & 0.1185 & 0.1181 & 0.1189 \\
	$1/\abar(m^{}_{Z_1})$ & $128.75\pm 0.09$ 
	& 128.75 & 128.75 & 128.75 & 128.73 & 128.75 & 128.75 \\
	$T_{\rm new}$ & & ----- & 0.0   & 0.0 & 0.0 & 0.0 & 0.0 \\
	$\xibar $ & & ----- 
	& 0.0001 & 0.0002 & $-0.0003$ & 0.0001 & 0.0016 \\
	$g_E^2/c_\chi^2 m_{Z_E}^2$& 
	& ----- & 0.279 & 1.771 & $-0.646$ & 0.668 & ----- \\
	\hline
	\end{tabular}
	\end{center}
	\caption{Summary of measurements for the low-energy 
	neutral current experiments~\cite{chm}. 
	The best fits are found by using all the electroweak 
	data of Table~2 and those in this Table. 
	The results for the $\chi, \psi, \eta$ and $\nu$ models are 
	obtained by setting $\delta = 0$. 
	The symbol $\eta^*$ denotes the leptophobic $\eta$-model 
	where $\delta$ is taken to be $\delta = 1/3$. 
	}
	\label{table:low_energy}
	\end{table}
	}
\def\llenc_table_old{
	\begin{table}[tp]
	\begin{center}
	\begin{tabular}{|r|c|r|r|r|r|r|r|}
	\hline
	 \multicolumn{2}{|c}{}& \multicolumn{6}{|c|}
	{pull = $\frac{\langle {\rm data} \rangle - {\rm SM}}
	{\langle {\rm error} \rangle}$} \\ 
\cline{3-8}
	 \multicolumn{2}{|c|}{}& SM & $\chi$ & $\psi$ & $\eta$ 
	& $\nu$ & $\eta^*$ 	\\ \hline
	\multicolumn{2}{|c|}{LENC experiments}&&&&&& \\ \hline
	$A_{\rm SLAC}$ & $0.80 \pm 0.058$ & 0.9 & 0.9 & 1.1 
		& 0.8 & 0.3 & 1.1 \\ 
	$A_{\rm CERN}$ & $-1.57 \pm 0.38$ &$-0.4$ & $-0.4$ & $-0.4$ 
		& $-0.5$ & $-0.6$ & $-0.4$ \\ 
	$A_{\rm Bates}$ & $-0.137 \pm 0.033$ & 0.5 & 0.4 & 0.4 
		& 0.4 & 0.4 & 0.4 \\ 
	$A_{\rm Mainz}$ & $-0.94 \pm 0.19$ & $-0.3$ & $-0.3$ & $-0.4$ 
		& 0.0 & 0.1 & $-0.4$ \\ 
	$Q_W(^{133}_{55}C_s)$ & $-72.08 \pm 0.92$ & 1.0 & 0.0 & 0.1 
		& 0.0 & 0.1 & 0.1 \\ 
	$K_{FH}$  & $0.3247 \pm 0.0040$ & $-1.1$ & $-1.1$ & $-1.1$ 
		& $-1.1$ & $-1.2$ & $-1.1$ \\ 
	$K_{CCFR}$  & $0.5820 \pm 0.0049$ & 0.6 & 0.6 & 0.6 
		& 0.6 & 0.4 & 0.6 \\ 
	$g_{LL}^{\nu_\mu e}$  & $-0.269 \pm 0.011$ & 0.4 & 0.1 & 0.3 
		& 0.2 & 0.0 & 0.4 \\ 
	$g_{LR}^{\nu_\mu e}$  & $0.234 \pm 0.011$ & 0.1& 0.0 & 0.2 
		& $-0.2$ & 0.3 & 0.1\\ 
	\hline
	\multicolumn{2}{|c|}{total $\chi^2$ and d.o.f.}
	&&&&&& \\ \hline
	$\chi^2_{\rm min}$ & & 4.1 & 2.8 & 3.2 & 2.7 & 2.3 & 3.2\\
	d.o.f. & & 9 & 6 & 6 & 6 & 6 & 7 \\ 
	\hline \hline
	\multicolumn{2}{|c|}{best fit} &&&&&& \\ \hline
	$m^{}_t$ & $175.6 \pm 5.5$ & 175.6 & 175.6 & 175.6 
		& 175.6 & 175.6 & 175.6 \\
	$1/\alpha (m^{}_Z)$ & $128.75\pm 0.09$ & 128.76 & 128.75
		& 128.75 & 128.75 & 128.75 & 128.75 \\
	$T_{\rm new}$ & & ----- & 0.014   & 0.005 & 0.0 & 0.0 & 0.004\\
	$\bar{\xi}$   & & ----- & 0.00010 & 0.010 & $-0.014$ 
		& $-0.0484$ & $-0.0064$ \\
	$g_E^2/c_\chi^2 m_{Z_E}^2$& 
	& ----- & 0.242 & 0.582 & 2.25 
		& 0.779 & ----- \\
	\hline
	\end{tabular}
	\caption{}
	\end{center}
	\end{table}
	}
\def\cvca_tab{	
	\begin{table}[t]
	\begin{center}
	\begin{tabular}{|l|c|c|}
	\hline
	& ${\it C_{fV}}$ & ${\it C_{fA}}$  \\ \hline
	$u$     & 3.1166 + 0.0030$x_s$ & 3.1351 + 0.0040$x_s$ \\ \hline
	$d = s$ & 3.1166 + 0.0030$x_s$ & 3.0981 + 0.0021$x_s$ \\ \hline
	$c$     & 3.1167 + 0.0030$x_s$ & 3.1343 + 0.0041$x_s$ \\ \hline
	$b$     & 3.1185 + 0.0030$x_s$ & 3.0776 + 0.0030$x_s$ \\ \hline
        $\nu$   & 1 & 1 \\ \hline
	$e=\mu$ & 1 & 1 \\ \hline
	$\tau$  & 1 & 0.9977 \\ \hline
	\end{tabular}
	\end{center}
	\caption{{\small Numerical values of factors $C_{fV}, 
	C_{fA}$ for quarks and leptons used in 
	eq.~(\ref{eq:partial_width}). 
	The finite mass corrections and the final state QCD corrections 
	for quarks are taken into accounted.  }}
	\vsp{0.3}
	\label{tab:cvca}
	\end{table}
	}
\def\zpole_result{
	\begin{table}[t]
	\caption{Summary of constraints on $T_{\rm new}$ 
	and $\xibar$ in $Z_\chi$, $Z_\psi$, $Z_\eta$ and $Z_\nu$ 
	models from $Z$-pole experiments and $m_W$ measurements. 
	The result of the leptophobic model, $Z_\eta$ with 
	$\delta = 1/3$, is also shown. 
	} 
	\begin{center}
	\begin{tabular}{cccccc}
	\\ \hline \hline
	model & $\delta$ & $T_{\rm new}$ & $\xibar$ & $\rho_{\rm corr}$ 
		& $\chi^2_{\rm min}/{\rm d.o.f.}$ 
	\\ \hline
	$\chi$ & 0 & $-0.040 \pm 0.115$ & $\hph 0.00017 \pm 0.00046$ 
		& $\hph 0.28$ & 16.5/12
\\
$\psi$ & 0& $-0.043 \pm 0.113$ & $\hph 0.00020 \pm 0.00050$ 
	& $\hph 0.20$ & 16.5/12
\\
$\eta$ & 0& $-0.053 \pm 0.111$ & $-0.00014 \pm 0.00108$ 
	& $\hph 0.09$ & 16.7/12
\\
$\nu$ & 0 & $-0.041 \pm 0.114$ & $\hph 0.00017 \pm 0.00042$ 
	& $\hph 0.22$ & 16.5/12
\\ \hline 
$\eta$ & 1/3& $-0.049 \pm 0.111$ & $\hph 0.00269 \pm 0.00309$ 
	& $\hph0.03$ & 15.9/12
\\ \hline \hline
\end{tabular}
\end{center}
\label{summary_e6}
\end{table}
}
\def\e6_extra{
\begin{table}[t]
\caption{
Charge assignment for the extra weak doublets in the minimal 
$E_6$ model and the $\eta_{\rm BKM}$ model of ref.~\cite{eta_model}. 
The symbol $a_i (-a_i)$ for $i=\chi,\psi,\eta,\nu$ are 
the ${\rm U(1)'}$ charge of $L$ or $H_d$ ($H_u$). 
}
\begin{center}
\begin{tabular}{ccccccc}  \hline \hline \\ [-0.4cm]
\vsp{0.2}
	& field & $Y$ & $2\sqrt{6}Q_\chi$ 
	& $\sqrt{72/5}Q_\psi$ & $Q_\eta$ 
	& $Q_\nu$ \\ \hline \\
\vsp{0.2}
minimal model & 
$H'$   & $-\frac{1}{2}$& $\hph a_\chi$ & $\hph a_\psi$ 
	& $\hph a_\eta$	& $\hph a_\nu$ 
\vsp{0.2} \\
&
$\ov{H}'$&$+\frac{1}{2}$& $-a_\chi$ & $-a_\psi$ & $-a_\eta$
	& $-a_\nu$\\
\hline 
$\eta_{\rm BKM}$ model~\cite{eta_model} & 
$H'_1$   & $-\frac{1}{2}$&  & & $\hph 1$ & 
\vsp{0.2} \\
&
$\ov{H_1}$& $+\frac{1}{2}$&  & & $-1$ & 
\vsp{0.2} \\
&
$H'_2$   & $-\frac{1}{2}$&  & & $\hph 1$ & 
\vsp{0.2} \\
&
$\ov{H_2}$& $+\frac{1}{2}$&  & & $-1$ & 
\vsp{0.2} \\
&
$D'$   & $-\frac{1}{3}$&  & & $\hph \frac{2}{3}$ & 
\vsp{0.2} \\
&
$\ov{D'}$& $+\frac{1}{3}$&  & & $-\frac{2}{3}$ & 
\\ \hline \hline
\end{tabular} 
\end{center}
\label{table:extra_higgs}
\end{table}
}
\def\coeff_rge{
\begin{table}[t]
\caption{Coefficients of the 1-loop $\beta$-functions 
	for the gauge couplings in the MSSM, the minimal 
	$E_6$ models and the $\eta_{\rm BKM}$ model~\cite{eta_model}.  
	The model $\chi(16)$ has three generations of ${\bf 16}$ 
	and a pair ${\bf 2}+{\bf\ov{2}}$. 
	The model $\chi(27)$ and $\psi, \eta,\nu$ have three 
	generations of ${\bf 27}$ and a pair ${\bf 2}+{\bf\ov{2}}$. 
	}
\begin{center}
\begin{tabular}{cccccccc}
\\ \hline \hline 
	& MSSM & $\chi(16)$ & $\chi(27)$ & $\psi$ & $\eta$ & $\nu$ 
	& $\eta_{\rm BKM}$~\cite{eta_model}
\\ \hline 
\vsp{0.05}\\
$b_1$	& $\frac{33}{5}$ & $\hph \frac{33}{5}$ 
	& $\frac{48}{5}$ & $\frac{48}{5}$ 
	& $\frac{48}{5}$ & $\frac{48}{5}$ 
	& $\frac{53}{5}$ \\
\vsp{0.05}\\
$b_2$	& 1    & \hph 1 & 4 & 4 & 4 & 4 & \hph 5 \\
\vsp{0.05}\\
$b_3$	& $-3$ & $-3$ & 0 & 0 & 0 & 0 & \hph 1 \\
\vsp{0.05}\\
$b_E$	& ---  & $6+ \frac{a^2}{10}$ & $9+ \frac{a^2}{10}$ 
	& $9+ \frac{a^2}{6}$ & $9+ \frac{12}{5}a^2$ 
	& $9+ \frac{12}{5}a^2$ & $\frac{77}{5}$ \\
\vsp{0.05}\\
$b_{1E}$& ---  & $-\sqrt{\frac{3}{50}}a$ & $-\sqrt{\frac{3}{50}}a$ 
	& $-\sqrt{\frac{1}{10}}a$ & $-\frac{6}{5}a$ 
	& $-\frac{6}{5}a$ & $-\frac{16}{5}$
\\ \hline \hline 
\end{tabular} 
\end{center}
\label{table:coeff_rge}
\end{table}
}
\def\ge_value{
\begin{table}[t]
\caption{Predictions for $g_E$ and $\delta$ at $\mu=m_{Z_1}$ 
	in the minimal models 
	and the $\eta_{\rm BKM}$ model~\cite{eta_model}. 
	The ${\rm U(1)}_Y$ gauge coupling $g_Y$ is fixed as 
	$g_Y = 0.36$. }
\begin{center}
\begin{tabular}{ccccc}
\\ \hline \hline 
model & $a$ & $g_E$ & $g_E/g_Y$ & $\delta$ \\
\hline 
$\chi(16)$ & $\hph 3$ & $\hph 0.353$& $0.989$ & $\hph 0.066$ \\
	   & $-2$ & $\hph 0.361$ & $1.010$ &$-0.044$ \\
\\
$\chi(27)$ & $\hph 3$ & $\hph 0.353$ & $0.989$ &$\hph 0.066$ \\
	   & $-2$ & $\hph 0.361$ & $1.010$ & $-0.044$ \\
\\
$\psi$ & $\hph 1$ & $\hph 0.364$ & $1.020$ &$\hph 0.028$ \\
	& $\hph 2$& $\hph 0.356$ & $0.999$ &$\hph 0.056$ \\
	&$-2$& $\hph 0.356$ & $0.999$ & $-0.056$ \\
\\
$\eta$ &$\hph 1/6$ & $\hph 0.366$ & $1.025$ &$\hph 0.018$ \\
	&$-2/3$ & $\hph 0.351$ & $0.982$ &$-0.071$ \\
\\
$\nu$ & $\hph \sqrt{1/6}$ & $\hph 0.361$& $1.010$ & $\hph 0.044$ \\
	&$-\sqrt{3/8}$ & $\hph 0.353$ & $0.989$ & $-0.066$ 
\\ \hline 
$\eta_{\rm BKM}$~\cite{eta_model} 
	& ---  & $\hph 0.308$ & $0.862$ &$\hph 0.286$ 
\\ \hline \hline 
\end{tabular}
\end{center}
\label{table:ge_delta}
\end{table}
}
\def\tnb_zeta{
\begin{table}[t]
\caption{Predictions for the effective $Z$-$Z'$ mixing parameter 
	$\zeta$ in the minimal $\chi,\psi,\eta,\nu$ and the 
	$\eta_{\rm BKM}$ model for $x^2 = 0$ and $0.5$, 
	and $\tan\beta = 2$ and $30$. 
	}
\begin{center}
\begin{tabular}{c|c|c|c|c|c} \hline \hline
\multicolumn{2}{c|}{}& \multicolumn{2}{c|}{$x^2 = 0$} 
	& \multicolumn{2}{c}{$x^2 = 0.5$} 
\\ \cline{3-6}
\multicolumn{2}{c|}{}& \multicolumn{2}{c|}{$\tan\beta$}
	& \multicolumn{2}{c}{$\tan\beta$} 
\\ \hline
& $a$ & 2 & 30 & 2 & 30 
\\ \hline
$\chi$ & $\hph 3$ & \multicolumn{2}{c}{$-0.88$} 
	& \multicolumn{2}{c}{$0.14$} 
\\
     & $-2$ & \multicolumn{4}{c}{$-0.77$}
\\ \hline
$\psi$ &$\hph 1$ & $\hph 0.60$ & $\hph 1.02$ & $\hph 0.55$ 
	& $\hph 0.76$
\\
       &$\hph 2$ & $\hph 0.58$ & $\hph 1.00$ & $\hph 0.79$ 
	& $\hph 1.00$
\\
       & $-2$& $\hph 0.69$ & $\hph 1.11$ & $-0.16$ 
	& $\hph 0.06$
\\ \hline
$\eta$ & $\hph 1/6$ & $-1.02$ & $-1.35$ & $-0.35$ & $-0.52$ 
\\ 
& $-2/3$ & $-0.93$ & $-1.26$ & $-1.11$ & $-1.26$ 
\\ \hline
$\nu$ & $\hph \sqrt{1/6}$ & $\hph 0.36$ & $\hph 0.77$ & $\hph 0.57$ 
	& $\hph 0.77$ 
\\ 
 & $-\sqrt{3/8}$ & $\hph 0.47$ & $\hph 0.88$ & $-0.34$ 
	& $-0.14$ 
\\ \hline
$\eta_{\rm BKM}$ & --- & $-1.29$ & $-1.62$ & $-1.79$ & $-1.95$
\\ \hline \hline
\end{tabular}
\end{center}
\label{table:zetasummary}
\end{table}
}
\def\masszeta{
\begin{table}[t]
\caption{Summary of the 95\% CL lower bound of $m_{Z_2}$ 
	(GeV) which corresponds to the predicted $\zeta$ in 
	Table~\ref{table:zetasummary}. 
	}
\begin{center}
\begin{tabular}{c|c|r|r|r|r} \hline \hline
\multicolumn{2}{c|}{}& \multicolumn{2}{c|}{$x^2 = 0$} 
	& \multicolumn{2}{c}{$x^2 = 0.5$} 
\\ \cline{3-6}
\multicolumn{2}{c|}{}& \multicolumn{2}{c|}{$\tan\beta$}
	& \multicolumn{2}{c}{$\tan\beta$} 
\\ \hline
& $a$ &$~~~~~~2$ & $~~~~30$ & $~~~~~~2$  & $~~~~30$
\\ \hline
$\chi$ & $\hph 3$ & \multicolumn{2}{c}{$1330$} 
	& \multicolumn{2}{c}{$620$} 
\\
     & $-2$ & \multicolumn{4}{c}{$1230$}
\\ \hline
$\psi$ &$\hph 1$ & $1290$ & $1800$ & $1220$ & $1480$
\\
       &$\hph 2$ & $1250$ & $1750$ & $1510$ & $1750$
\\
       & $-2$& $1380$ & $1890$ & $520$ & $370$
\\ \hline
$\eta$ & $+1/6$ & $1330$ & $1690$ & $620$ & $790$ 
\\ 
& $-2/3$ & $1230$ & $1590$ & $1410$ & $1590$
\\ \hline
$\nu$ & $-\sqrt{3/8}$ & $1180$ & $1720$ & $800$ & $520$ 
\\ 
& $+\sqrt{1/6}$ & $1030$ & $1580$ & $1320$ & $1580$ 
\\ \hline
$\eta_{\rm BKM}$ & ---  & $1520$ & $1930$ & $2150$ & $2360$
\\ \hline \hline
\end{tabular}
\end{center}
\label{table:mass95_zeta}
\end{table}
}

\def\bound_lenc{
\begin{table}[t]
\caption{The 95\% CL lower bound of $m_{Z_2}$ (GeV) in the 
	$\chi, \psi, \eta$ and $\nu$ models ($\delta = 0$) 
	for $g_E=g_Y$ and $\mh = 100~{\rm GeV}$. 
	The results of previous study~\cite{cvetic_langacker_review} 
	and of recent direct search~\cite{direct_search} are 
	shown for comparison. }
\begin{center}
\begin{tabular}{c|cccc} \hline \hline 
	& $\chi$ & $\psi$ & $\eta$ & $\nu$ \\ \hline
Our results & 451 & 136 & 317 & 284 \\ 
Langacker \etal~\cite{cvetic_langacker_review} 
	& 330 & 170 & 220 & --- \\
direct search~\cite{direct_search} 
	& 595 & 590 & 620 & --- 
\\ \hline \hline 
\end{tabular}
\end{center}
\label{mzelenc}
\end{table}  
	}
\def\tnew_xi{
	\begin{figure}[t]
	\begin{center}
	\leavevmode\psfig{figure=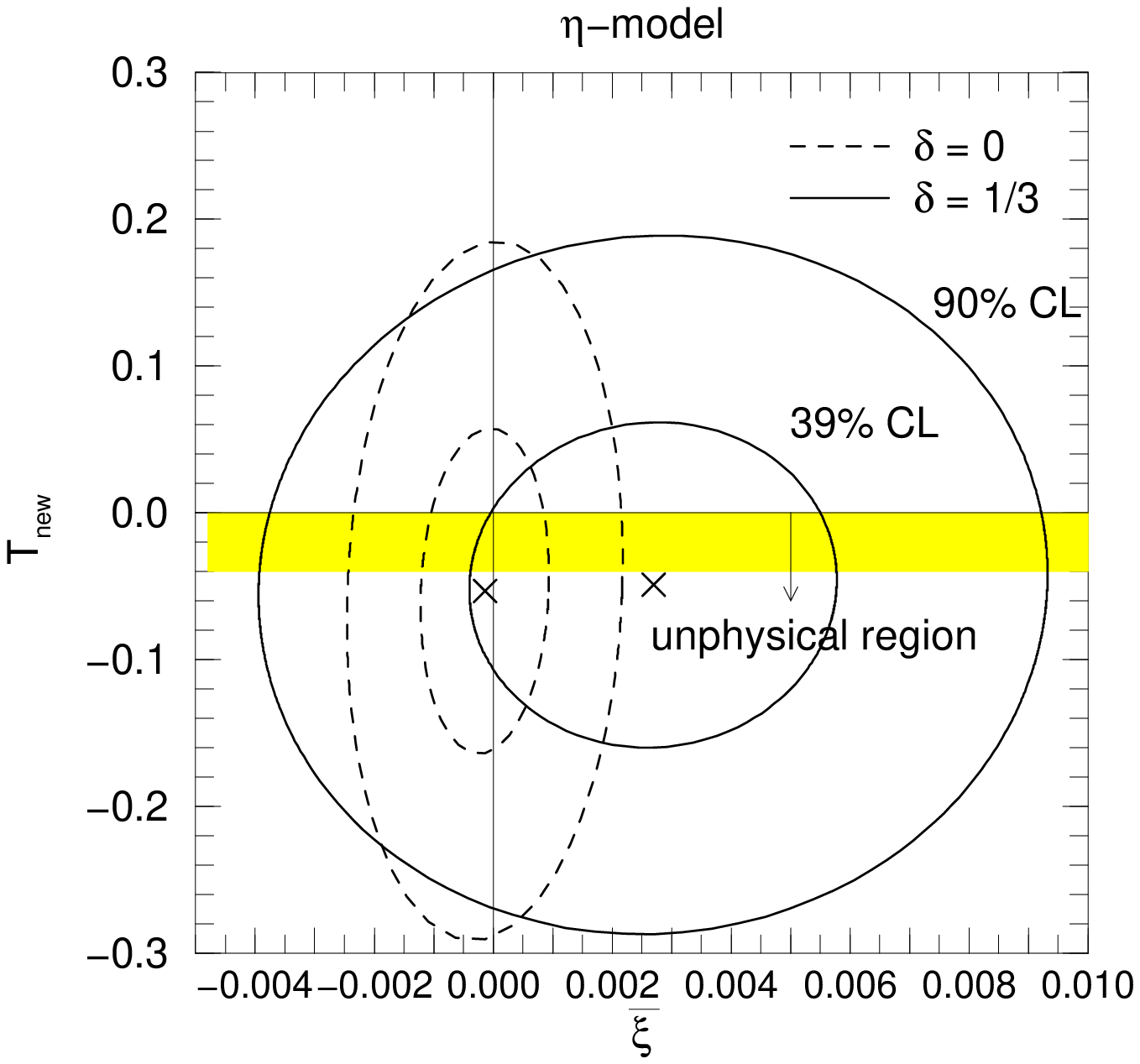,width=9cm}
	\end{center}
	\caption{The 1-$\sigma$ and 90\% CL allowed region on 
	the $(\xibar, T_{\rm new})$ plane in the $\eta$-model 
	with $\delta =0$ and $1/3$. 
	The shaded region ($T_{\rm new} < 0$)
	corresponds to the unphysical region. }
	\label{allowed_eta}
	\end{figure}
}
\def\fig_beta_delta{
	\begin{figure}[t]
	\begin{center}
	\leavevmode\psfig{figure=fig2.eps,width=9cm}
	\end{center}
	\caption{Contour plot of 
	$\Delta \chi^2 \equiv \chi_{\rm min}^2(\beta,\delta) 
	- \chi^2_{\rm min}({\rm SM})$ for $\mh = 100~{\rm GeV}$. 
	The mixing angle $\beta_E$ for the $\chi,\psi,\eta$ and 
	$\nu$ models are shown by vertical dotted lines. 
	The step of each contour is 0.2. 
	The ``$\times$'' marks on the plot show the specific models 
	listed in Table~\ref{table:ge_delta} in Sec.~5. }
	\label{chisq_distribution}
	\end{figure}
}
\def\mass_lenc{
	\begin{figure}[t]
	\begin{center}
	\leavevmode\psfig{figure=fig3.eps,width=9cm}
	\end{center}
	\caption{Contour plot of the 95\% CL lower mass 
	limit of the $Z_2$ boson obtained from the LENC experiments 
	for $g_E = g_Y$ and $\mh = 100~{\rm GeV}$. 
	The vertical dotted lines correspond to 
	the $\chi, \psi, \eta$ and $\nu$ models.
	The limits are given in the unit of GeV.
	The ``$\times$'' marks on the plot show the specific models 
	listed in Table~\ref{table:ge_delta} in Sec.~5. }
	\label{mass_distribution}
	\end{figure}
	}
\def\zprime_mass{
	\begin{figure}[t]
	\begin{center}
	\leavevmode\psfig{figure=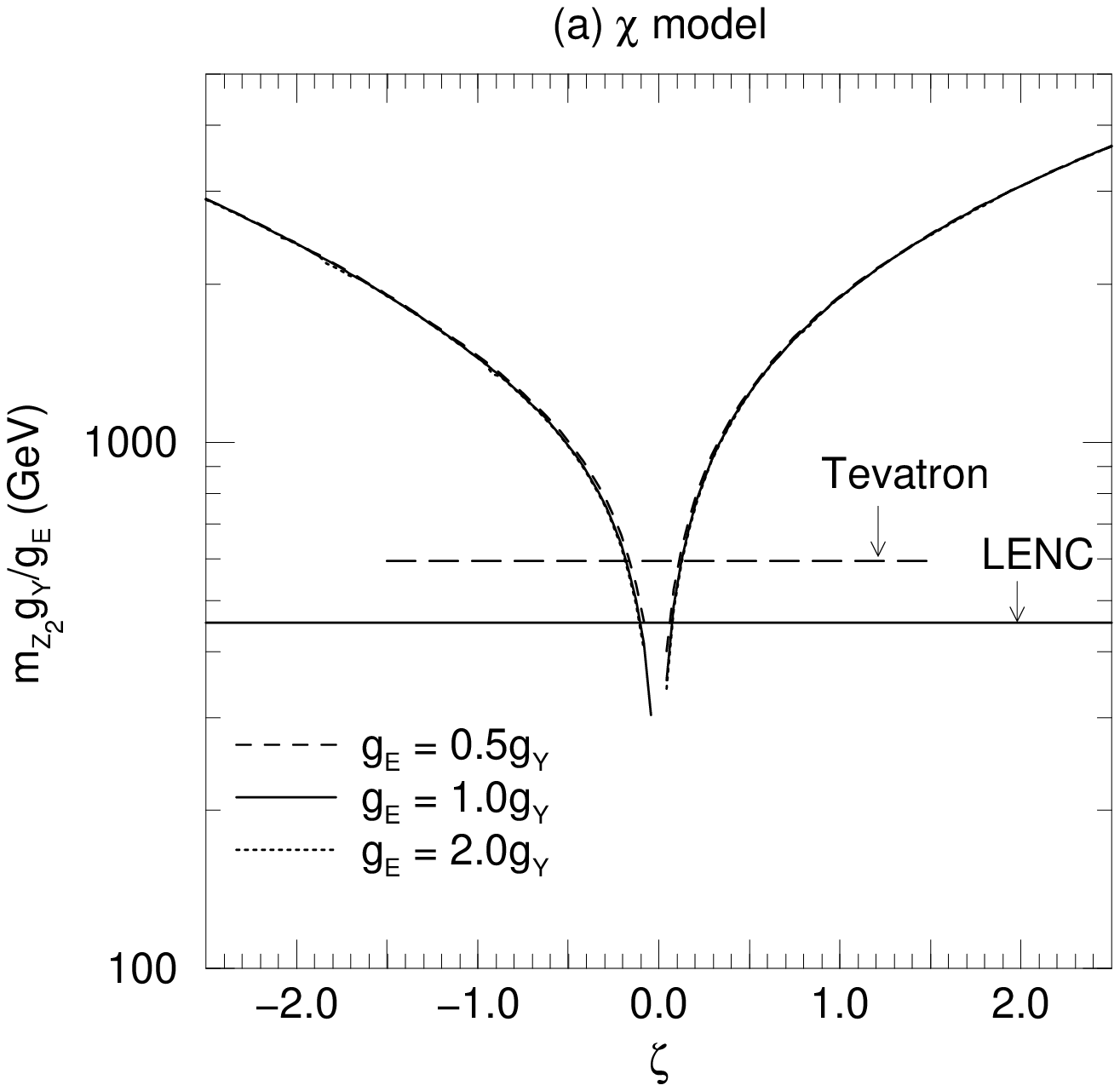,width=6cm}
	\leavevmode\psfig{figure=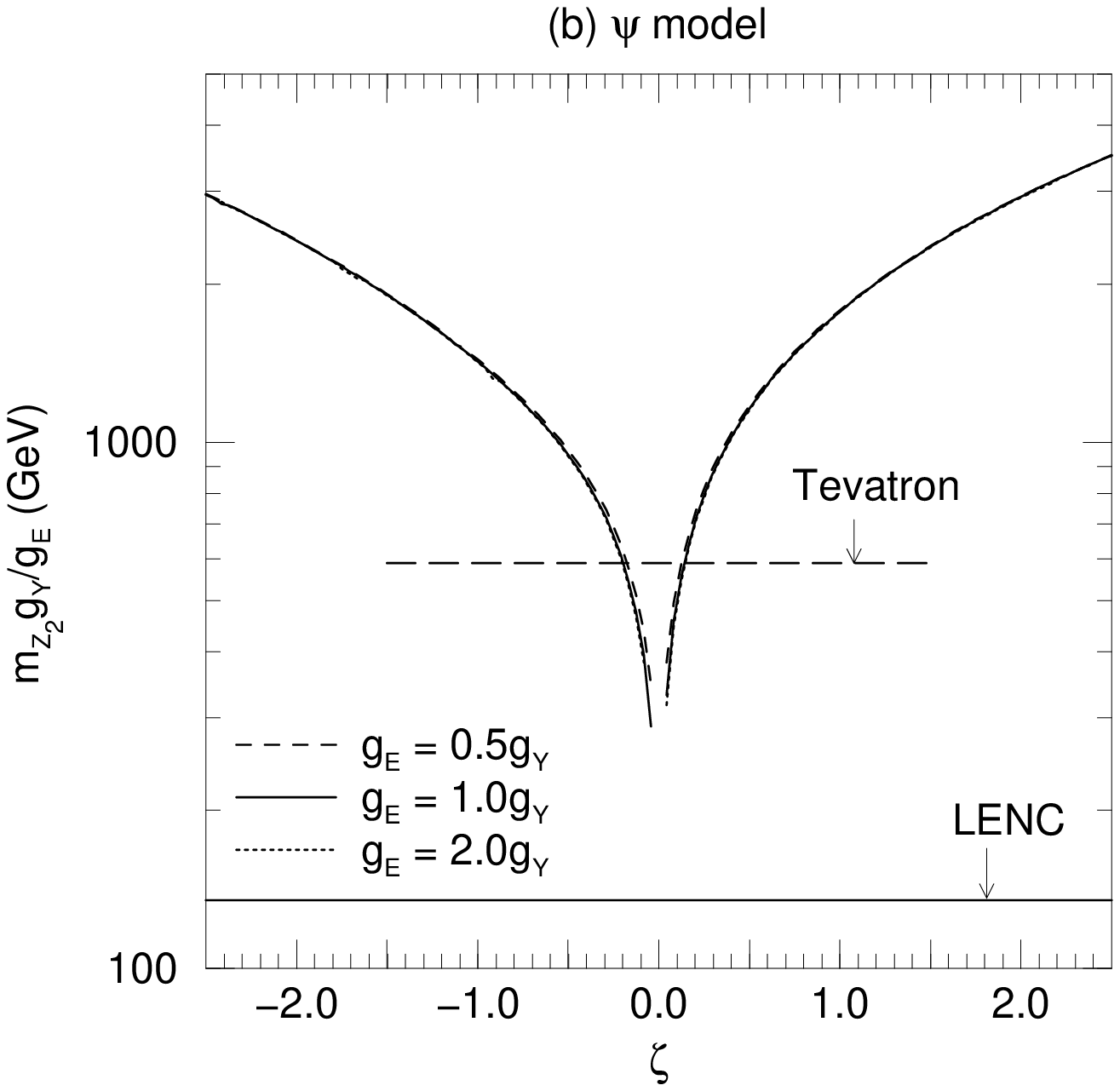,width=6cm}
	\leavevmode\psfig{figure=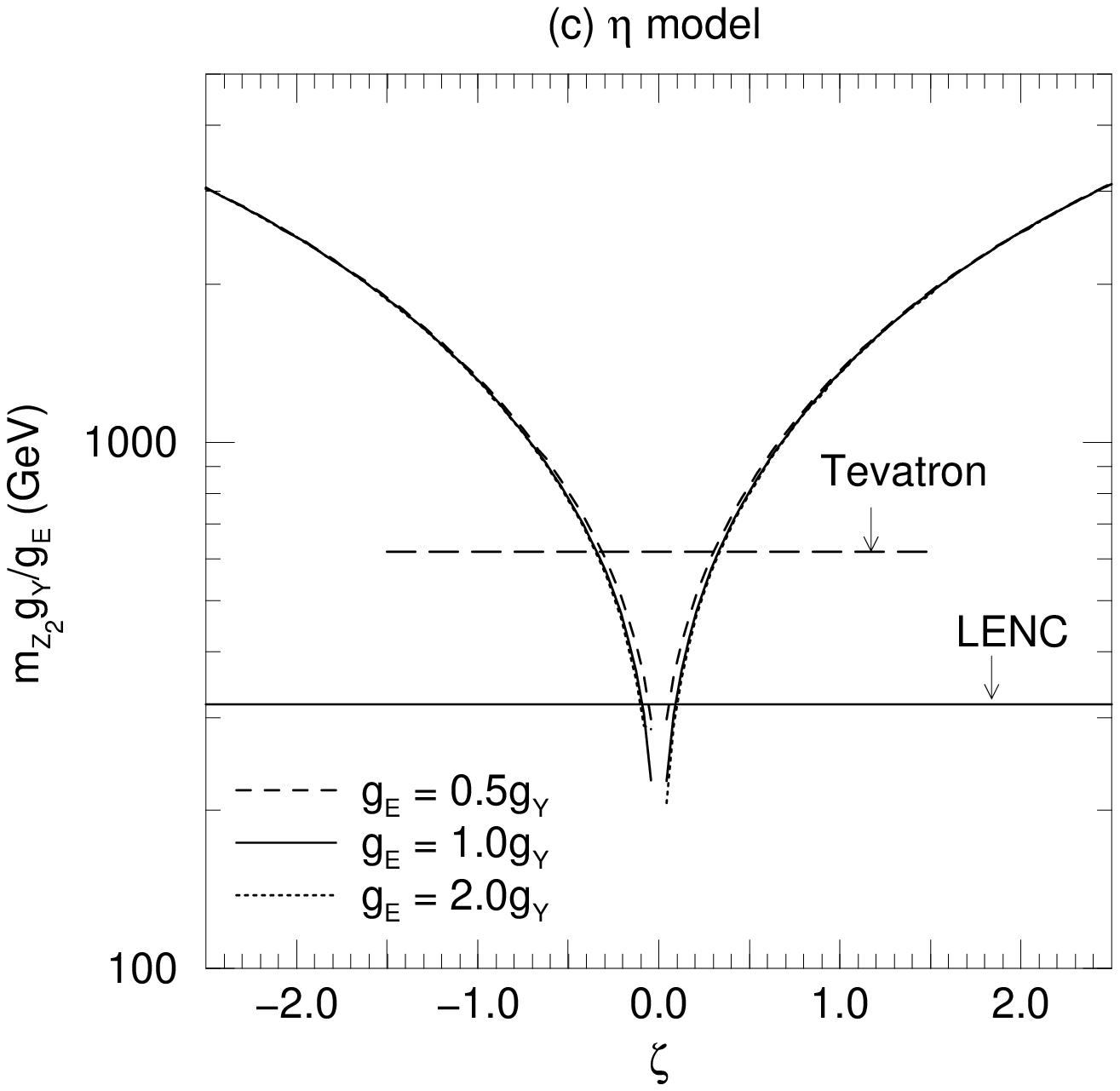,width=6cm}
	\leavevmode\psfig{figure=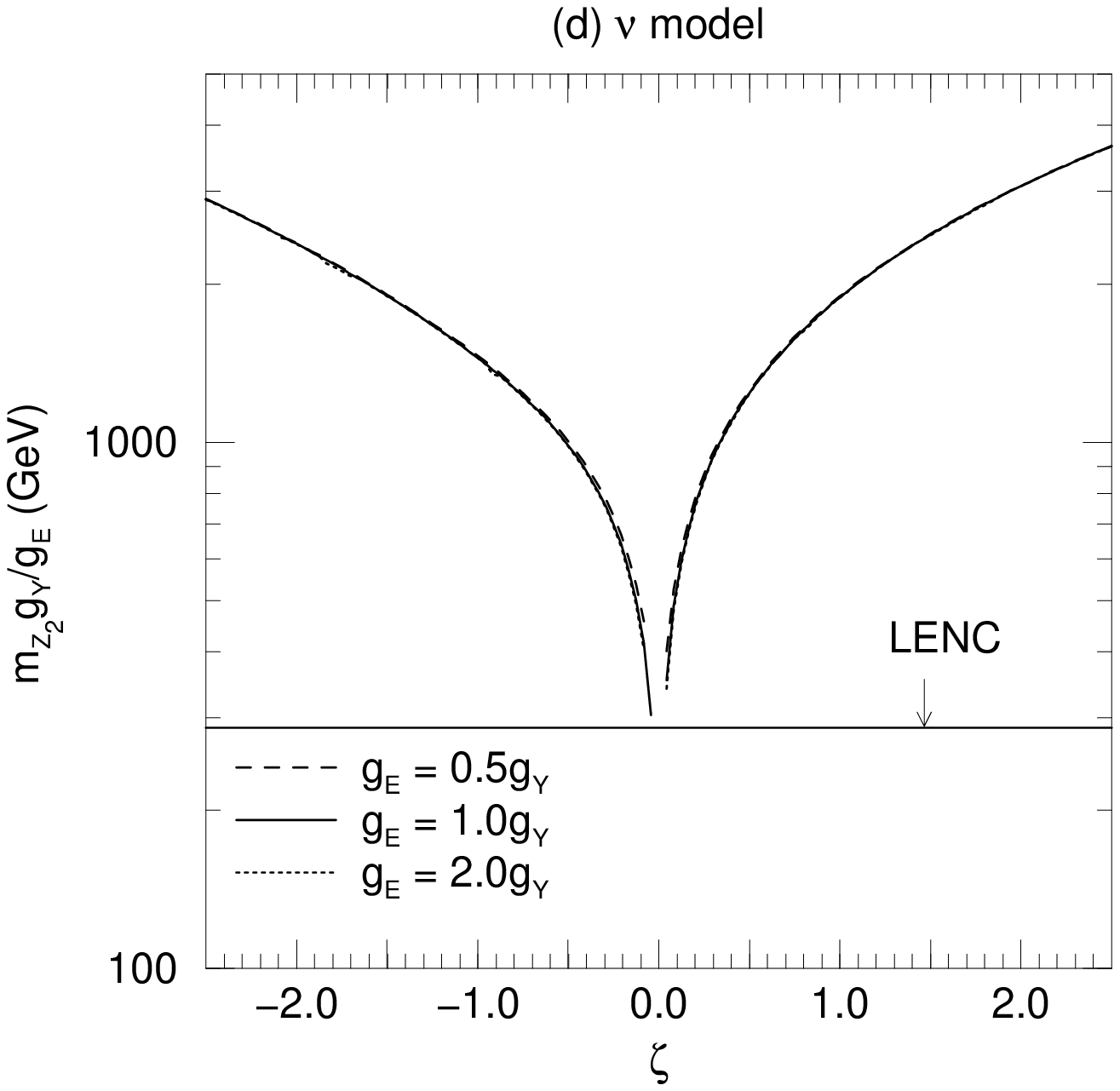,width=6cm}
	\end{center}
	\caption{The 95\% CL lower mass limit of $Z_2$ in the 
	$\chi$, $\psi$, $\eta$ and $\nu$ models 
	for $\mh = 100~{\rm GeV}$. 
	The $Z_2$ boson mass is normalized by $g_E/g_Y$. 
	Constraints from $Z$-pole experiments and LENC 
	experiments are separately shown. 
	The results of the direct search at Tevatron~\cite{direct_search}
	for the $\chi,\psi$ and $\eta$ models are also 
	shown.  
	}
	\label{mass_95cl}
	\end{figure}
}
\def\MPLA#1#2#3{Mod. Phys. Lett. {\bf A#1} (19#2) #3}
\def\PRD#1#2#3{Phys. Rev. {\bf D#1} (19#2) #3}
\def\NPB#1#2#3{Nucl. Phys. {\bf B#1} (19#2) #3}
\def\PTP#1#2#3{Prog. Theor. Phys. {\bf #1} (19#2) #3}
\def\ZPC#1#2#3{Z. Phys. {\bf C#1} (19#2) #3}
\def\EPJC#1#2#3{Eur. Phys. J. {\bf C#1} (19#2) #3}
\def\PLB#1#2#3{Phys. Lett. {\bf B#1} (19#2) #3}
\def\PRL#1#2#3{Phys. Rev. Lett. {\bf #1} (19#2) #3}
\def\PR#1#2#3{Phys. Rep. {\bf #1} (19#2) #3}
\makeatletter
\newtoks\@stequation

\def\subequations{\refstepcounter{equation}%
  \edef\@savedequation{\the\c@equation}%
  \@stequation=\expandafter{\theequation}
  \edef\@savedtheequation{\the\@stequation}
  \edef\oldtheequation{\theequation}%
  \setcounter{equation}{0}%
  \def\theequation{\oldtheequation\alph{equation}}}

\def\endsubequations{%
  \ifnum\c@equation < 2 \@warning{Only \the\c@equation\space subequation
    used in equation \@savedequation}\fi
  \setcounter{equation}{\@savedequation}%
  \@stequation=\expandafter{\@savedtheequation}%
  \edef\theequation{\the\@stequation}%
  \global\@ignoretrue}

\def\eqnarray{\stepcounter{equation}\let\@currentlabel\theequation
\global\@eqnswtrue\m@th
\global\@eqcnt\z@\tabskip\@centering\let\\\@eqncr
$$\halign to\displaywidth\bgroup\@eqnsel\hskip\@centering
     $\displaystyle\tabskip\z@{##}$&\global\@eqcnt\@ne
      \hfil$\;{##}\;$\hfil
     &\global\@eqcnt\tw@ $\displaystyle\tabskip\z@{##}$\hfil
   \tabskip\@centering&\llap{##}\tabskip\z@\cr}

\makeatother

\begin{document}
\thispagestyle{empty}
\vspace*{-15mm}
\baselineskip 10pt
\begin{flushright}
\begin{tabular}{l}
{\bf KEK-TH-576}\\
{\bf EPHOU-98-006}\\
{\bf hep-ph/9805448}\\
{\bf May 1998}
\end{tabular}
\end{flushright}
\baselineskip 24pt 
\vglue 10mm 
\begin{center}
{\Large\bf
$Z'$ bosons in supersymmetric $E_6$ models 
confront electroweak data
}
\vspace{5mm}

\baselineskip 18pt 
\def\thefootnote{\fnsymbol{footnote}}
\setcounter{footnote}{0}
{\bf
Gi-Chol Cho$^{1}$\footnote{Research Fellow of the Japan Society 
for the Promotion of Science}, 
Kaoru Hagiwara$^{1}$ and Yoshiaki Umeda$^{1,2}$ }
\vspace{3mm}

$^1${\it Theory Group, KEK, Tsukuba, Ibaraki 305-0801, Japan}\\
$^2${\it Department of Physics, Hokkaido University, Sapporo 
060-0810, Japan} 
\vspace{10mm}
\end{center}

\begin{center}
{\bf Abstract}\\[7mm]
\begin{minipage}{13.5cm}
\baselineskip 16pt
\noindent
We study constraints on additional $Z'$ bosons 
predicted in the supersymmetric (SUSY) $E_6$ models by using 
the updated results of electroweak experiments 
-- $Z$-pole experiments, $m_W$ measurements and 
low-energy neutral current (LENC) experiments. 
We find that the effects of $Z$-$Z'$ mixing are 
parametrized by 
(i) a tree-level contribution to the $T$-parameter,  
(ii) the effective $Z$-$Z'$ mass mixing angle $\xibar$. 
In addition, the effect of the direct exchange of the heavier 
mass eigenstate $Z_2$ in the LENC processes is parametrized
by (iii) a contact term $\contact$. 
We give the theoretical predictions for the observables 
in the electroweak experiments together with the standard 
model radiative corrections. 
Constraints on $T_{\rm new}$ and $\xibar$ 
from the $Z$-pole and $m_W$ experiments and those 
on $\contact$ from the LENC experiments 
are separately shown. 
Impacts of the kinetic mixing between 
the ${\rm U(1)}_Y$ and ${\rm U(1)'}$ gauge bosons on 
the $\chi^2$-analysis are studied. 
We show the 95\% CL lower mass limit of $Z_2$ as a function of 
the effective $Z$-$Z'$ mixing parameter $\zeta$, a combination 
of the mass and kinetic mixings. 
Theoretical prediction on $\zeta$ and $g_E$ is found for 
the $\chi, \psi, \eta$ and $\nu$ models by assuming 
the minimal particle content of the SUSY $E_6$ models. 
In a certain region of the parameter space, 
the $Z_2$ boson mass in the detectable range of LHC 
is still allowed. 
\end{minipage}
\end{center}
\baselineskip 18pt 
{\small 
\begin{flushleft}
\end{flushleft}
}

\newpage
\baselineskip 18pt 
\def\thefootnote{\arabic{footnote}}
\setcounter{footnote}{0}
%
%
%
%
\section{Introduction}
Although the minimal Standard Model (SM) agrees well with 
current electroweak experiments~\cite{lep_slc_97}, 
it is important to examine consequences of new physics 
models beyond the SM at current or future collider experiments. 
One of the simplest extensions of the SM is to introduce an 
additional U(1) gauge symmetry, ${\rm U(1)'}$, 
whose breaking scale is close to the electroweak scale. 
The ${\rm U(1)'}$ symmetry is predicted in a certain class 
of grand unified theories (GUTs) with gauge group whose rank 
is higher than that of the SM. 
In general, the additional ${\rm U(1)'}$ gauge boson 
$Z'$ can mix with the hypercharge ${\rm U(1)}_Y$ gauge boson 
through the kinetic term at above the electroweak scale, 
and also it can mix with the SM $Z$ boson after the 
electroweak symmetry is spontaneously broken.  
Through those mixings, the $Z'$ boson can affect the electroweak 
observables at the $Z$-pole and the $W$ boson mass $m_W$. 
Both the $Z$-$Z'$ mixing and the direct $Z'$ contribution 
can affect neutral current experiments off the $Z$-pole. 
The presence of an additional $Z'$ boson can be explored 
directly at $p \bar{p}$ collider experiments. 

The supersymmetric (SUSY) $E_6$ models are the promising candidates 
which predict an additional $Z'$ boson at the weak scale (for 
a review, see~\cite{hewett_rizzo}). 
The gauge group $E_6$ can arise from the perturbative heterotic 
string theory as a consequence of its compactification.  
In the $E_6$ models, the SM matter fields in each generation 
are embedded into its fundamental representation ${\bf 27}$ 
that also contains several exotic matter fields -- two SM singlets, 
a pair of weak doublets and color triplets. 
Because $E_6$ is a rank-six group, it can have two extra 
${\rm U(1)}$ factors besides the SM gauge group. 
A superposition of the two extra ${\rm U(1)}$ groups may 
survive as the ${\rm U(1)'}$ gauge symmetry at the GUT scale. 
The ${\rm U(1)'}$ symmetry may break spontaneously at the weak 
scale through the radiative corrections to the mass term of the 
SM singlet scalar field~\cite{radiative_u1prime}. 

In this paper, we study constraints on the $Z'$ bosons predicted 
in the SUSY $E_6$ models. 
Although there are several previous works~\cite{altarelli,
langacker_luo,chm,gherghetta,zprime_lep}, we would like to 
update their studies by using the recent results of 
electroweak experiments, and by allowing for an arbitrary 
kinetic mixing~\cite{holdom, eta_model,general_zzmixing} 
between the $Z'$ boson and the hypercharge $B$ boson. 
In our study, we use the results of $Z$-pole experiments 
at LEP1 and SLC, and the $m_W$ measurements at Tevatron and LEP2 
which were reported at the summer conferences in 
1997~\cite{lep_slc_97}. 
We also study the constraints from low-energy neutral current 
(LENC) experiments: lepton-quark, lepton-lepton scattering 
experiments and atomic parity violation measurements. 

We find that the lower mass limit of the heavier mass 
eigenstate $Z_2$ is obtained as a function of the effective $Z$-$Z'$ 
mixing term $\zeta$, which is a combination of the mass and kinetic 
mixings. 
In principle, $\zeta$ is calculable, together with the gauge 
coupling $g_E$, once the particle spectrum of the $E_6$ model 
is specified. 
We show the theoretical prediction for $\zeta$ and $g_E$ 
in the SUSY $E_6$ models by assuming the minimal particle 
content which satisfies the anomaly free condition and the 
gauge coupling unification.  
For those models, the electroweak data give stringent 
lower mass bound on the $Z_2$ boson. 

This paper is organized as follows. 
In the next section, we briefly review the additional $Z'$ 
boson in the SUSY $E_6$ models and the generic feature of 
$Z$-$Z'$ mixing in order to fix our notation. 
We show that the effects of $Z$-$Z'$ mixing and direct 
$Z'$ boson contribution are parametrized 
by the following three terms: 
(i) a tree-level contribution to the $T$ parameter~\cite{peskin_takeuchi}, 
$T_{\rm new}$, 
(ii) the effective $Z$-$Z'$ mass mixing angle $\xibar$ 
and 
(iii) a contact term $\contact$ which appears in the low-energy 
processes. 
In Sec.~3, we collect the latest results of electroweak experiments. 
There, the theoretical predictions for the electroweak 
observables are shown together with the SM radiative corrections. 
In Sec.~4, we show constraints on the $Z'$ bosons from the 
electroweak data. 
The presence of non-zero kinetic mixing between the 
${\rm U(1)}_Y$ and ${\rm U(1)'}$ gauge bosons 
modifies the couplings between the $Z'$ boson and the SM 
fermions. 
We discuss impacts of the kinetic mixing term 
on the $\chi^2$-analysis. 
The 95\% CL lower mass limit of the heavier mass eigenstate $Z_2$ 
is given as a function of the effective $Z$-$Z'$ mixing parameter 
$\zeta$. 
The $\zeta$-independent constraints from the low-energy 
experiments and those from the direct search experiments 
at Tevatron are also discussed. 
In Sec.~5, 
we find the theoretical prediction for $\zeta$ in some 
SUSY $E_6$ models ($\chi, \psi, \eta, \nu$) 
by assuming the minimal particle content.  
Stringent $Z_2$ boson mass bounds are found for most models.
Sec.~6 summarizes our findings. 
%
%
%
%
\section{$Z$-$Z'$ mixing in supersymmetric $E_6$ model}
\subsection{$Z'$ boson in supersymmetric $E_6$ model}
\clean 
Since the rank of $E_6$ is six, 
it has two ${\rm U(1)}$ factors besides the SM gauge 
group which arise from the following decompositions: 
\bea
	\begin{array}{rl}
	E_6 &\supset {\rm SO(10)} \times {\rm U(1)}_\psi  
\\
	&\supset {\rm SU(5)} \times {\rm U(1)}_\chi \times {\rm U(1)}_\psi. 
	\end{array}
\eea
\label{eq:e6_breaking}
\hsp{-0.3}
An additional $Z'$ boson in the electroweak scale 
can be parametrized as 
a linear combination of the ${\rm U(1)}_\psi$ gauge boson 
$Z_\psi$ and the ${\rm U(1)}_\chi$ gauge boson 
$Z_\chi$ as~\cite{PDG} 
\beq
Z' = Z_\chi \cos \beta_E + Z_\psi \sin \beta_E. 
\eeq
In this paper, we study the following four $Z'$ models 
in some detail: 
\bea
\begin{array}{|c||c|c|c|c|} \hline 
~~~\beta_E~~~ & ~~~0~~~ &~~~\pi/2~~~ & \tan^{-1}(-\sqrt{5/3}) 
& \tan^{-1}(\sqrt{15}) \\ \hline
{\rm model} & \chi      & \psi         & \eta      & \nu      
\\  \hline 
\end{array}
\eea
In the SUSY-$E_6$ models, each generation of the SM quarks 
and leptons is embedded into a {\bf 27} representation. 
In Table~1, we show all the matter fields 
contained in a {\bf 27} and their classification in SO(10) 
and SU(5). 
The ${\rm U(1)'}$ charge assignment on the matter fields 
for each model is also given in the same table. 
The normalization of the ${\rm U(1)'}$ charge follows 
that of the hypercharge. 
\esix_contents
Besides the SM quarks and leptons, there are two SM singlets 
$\nu^c$ and $S$, a pair of weak doublets $H_u$ and $H_d$, 
a pair of color triplets $D$ and $\ov{D}$ in each generation. 
The $\eta$-model arises when $E_6$ breaks into a rank-5 group 
directly in a specific compactification of the heterotic 
string theory~\cite{eta_ellis}. 
In the $\nu$-model, the right-handed neutrinos $\nu^c$ are 
gauge singlet~\cite{nu_model} and can have large Majorana 
masses to realize the see-saw mechanism~\cite{see-saw}. 

The ${\rm U(1)'}$ symmetry breaking occurs if the scalar 
component of the SM singlet field develops the vacuum 
expectation value (VEV). 
It can be achieved at near the weak scale via radiative 
corrections to the mass term of the SM singlet scalar 
field. 
For example, the terms $SD\ov{D}$ and $S H_u H_d$ appear 
in the ${\rm SU(3)}_C \times {\rm SU(2)}_L \times 
{\rm U(1)}_Y \times {\rm U(1)'}$ invariant superpotential. 
If the Yukawa couplings of the $SD\ov{D}$ term and/or 
$S H_u H_d$ term are $O(1)$, the squared mass of the 
scalar component of $S$ can become negative at the weak 
scale through the renormalization group equations (RGEs) 
with an appropriate boundary condition at the GUT scale. 
Recent studies of the radiative ${\rm U(1)'}$ symmetry 
breaking can be found, \eg, in ref.~\cite{radiative_u1prime}. 

Several problems may arise in the $E_6$ models from view 
of low-energy phenomenology~\cite{hewett_rizzo}.  
For example, the scalar components of extra colored triplets 
$D, \ov{D}$ in ${\bf 27}$ could mediate an instant proton 
decay. 
It should be forbidden by imposing a certain discrete symmetry 
on the general 
${\rm SU(3)}_C \times {\rm SU(2)}_L \times {\rm U(1)}_Y \times
{\rm U(1)'}$ invariant superpotential. 
Except for the $\nu$-model~\cite{nu_model}, the large Majorana 
mass of $\nu^c$ is forbidden by the ${\rm U(1)'}$ gauge 
symmetry, and the fine-tuning is needed to make the Dirac neutrino 
mass consistent with the observation. 
Further discussions can be found in ref.~\cite{hewett_rizzo}.  
In the following, we 
assume that these requirements are satisfied by an unknown 
mechanism. 
Moreover we assume that all the super-partners of the SM 
particles and the exotic matters do not affect the radiative 
corrections to the electroweak observables significantly,  
\ie, they are assumed to be heavy enough to decouple from 
the weak boson mass scale. 

\subsection{Phenomenological consequences of $Z$-$Z'$ mixing}
If the SM Higgs field carries a non-trivial ${\rm U(1)'}$ 
charge, its VEV induces the $Z$-$Z'$ mass mixing. 
On the other hand, the kinetic mixing between the hypercharge 
gauge boson $B$ and the ${\rm U(1)'}$ gauge boson $Z'$ 
can occur through the quantum effects below the GUT scale. 
After the electroweak symmetry is broken, the effective 
Lagrangian for the neutral gauge bosons in the 
${\rm SU(2)}_L \times {\rm U(1)}_Y \times {\rm U(1)'}$ 
theory is given by~\cite{eta_model} 
\bea
{\cal L}_{gauge} 
	&=&  -\frac{1}{4}Z^{\mu\nu}Z_{\mu\nu}
            -\frac{1}{4}Z'^{\mu\nu}Z'_{\mu\nu} 
	    -\frac{\sin \chi}{2}B^{\mu\nu}Z'_{\mu\nu}
	    -\frac{1}{4}A^{0\mu\nu}A^{0}_{\mu\nu} 
\nonumber \\ 
	& & + m^2_{ZZ'} Z^{\mu}Z'_{\mu}
	    +\frac{1}{2} m^2_Z Z^{\mu}Z_{\mu}
	    +\frac{1}{2} m^2_{Z'} Z'^\mu Z'_{\mu}, 
\label{eq:l_gauge}
\eea 
where $F^{\mu\nu} (F=Z,Z',A^0,B)$ represents the gauge field strength. 
The $Z$-$Z'$ mass mixing and the kinetic mixing are characterized 
by $m^2_{ZZ'}$ and $\sin \chi$, respectively. 
In this basis, the interaction Lagrangian for the neutral current 
process is given as 
\begin{eqnarray}
{\cal L}_{NC} &=& -\sum_{f,\, \alpha}  \left\{ 
	\; e \, Q_{f^{}_{\alpha}} \overline{f^{}_{\alpha}}
	\gamma^{\mu}f^{}_{\alpha} A^0_{\mu} +
	g^{}_Z \overline{f^{}_{\alpha}} \gamma^{\mu}
	\left( I^3_{f_L} - Q_{f^{}_{\alpha}} \sin^2\theta_W \right)
	f^{}_{\alpha} Z^{}_{\mu} \right. \nonumber \\ 
	& & \left. + g^{}_E Q^{f^{}_{\alpha}}_E 
	\overline{f^{}_{\alpha}}\gamma^{\mu}f^{}_{\alpha} 
	Z'_{\mu} \right\}, 
\label{eq:neutraC}
\end{eqnarray}
where $g_Z = g/\cos\theta_W = g_Y/\sin\theta_W$. 
The ${\rm U(1)'}$ gauge coupling constant is denoted by $g_E$ 
in the hypercharge normalization. 
The symbol $f_\alpha$ denotes the quarks or leptons with 
the chirality $\alpha$ ($\alpha = L$ or $R$). 
The third component of the weak isospin, the electric charge 
and the ${\rm U(1)'}$ charge of $f_\alpha$ are given by 
$I^3_{f_\alpha}$, $Q_{f_\alpha}$ and $Q_E^{f_\alpha}$, respectively. 
The ${\rm U(1)'}$ charge of the quarks and leptons listed in Table~1 
should be read as 
\bea
	\left.
	\begin{array}{l}
Q_E^Q = Q_E^{u_L} = Q_E^{d_L},~~~Q_E^L = Q_E^{\nu_L} = Q_E^{e_L}, 
\\ 
Q_E^{f^c} = -Q_E^{f_R} ~~~(f=e,u,d),
	\end{array}
	\right \}. 
\label{eq:charge_rule}
\eea
The mass eigenstates $(Z_1,Z_2,A)$ is obtained by 
the following transformation; 
\begin{equation}
\left( \begin{array}{c}  Z \\  Z' \\  A^0  
\end{array}\right) 
= 
\left(
	\begin{array}{ccc}
\cos \xi + \sin \xi \sin \theta_W \tan \chi &
-\sin \xi + \cos \xi \sin\theta_W \tan \chi & 0 \\
\sin \xi / \cos \chi & \cos \xi / \cos \chi & 0 \\
-\sin\xi \cos \theta_W \tan \chi & 
- \cos \xi \cos \theta_W \tan \chi & 1   
	\end{array}
\right) 
\left( \begin{array}{c} {Z_1} \\ 
{Z_2} \\ {A} \end{array}\right). 
\end{equation}
Here the mixing angle $\xi$ is given by 
\begin{equation}
\tan 2\xi = \frac{-2c^{}_{\chi}(m^2_{ZZ'}+s^{}_W s^{}_{\chi}
            m^2_Z)}{m^2_{Z'} - (c^2_{\chi}-s^2_W s^2_{\chi})m^2_Z+
            2s^{}_W s^{}_{\chi} m^2_{ZZ'}}~, 
\label{eq:angle_xi}
\end{equation}
with the short-hand notation, 
$c_\chi =  \cos\chi$, $s_\chi = \sin\chi$ and $s_W = \sin\theta_W$. 
The physical masses $m_{Z_1}$ and $m_{Z_2}$ ($m_{Z_1} < m_{Z_2}$) 
are given as follows; 
\bsub
\bea
m_{Z_{1}}^2 
	 &=& m_Z^2 (c_\xi + s_\xi s_W t_\chi)^2 
	+ m_{Z'}^2 \biggl( \frac{s_\xi}{c_\chi} \biggr)^2
	+ 2 m^2_{ZZ'} \frac{s_\xi}{c_\chi} (c_\xi + s_\xi s_W t_\chi),  
\label{eq:light_Z1}
\\
m_{Z_{2}}^2 
	 &=& m_Z^2 (c_\xi s_W t_\chi - s_\xi)^2 
	+ m_{Z'}^2 \biggl( \frac{c_\xi}{c_\chi} \biggr)^2
	+ 2 m^2_{ZZ'} \frac{c_\xi}{c_\chi} (c_\xi s_W t_\chi -s_\xi), 
\eea
\esub
where $c_\xi = \cos\xi$, $s_\xi = \sin\xi$ and $t_\chi = \tan\chi$. 
The lighter mass eigenstate $Z_1$ should be identified with 
the observed $Z$ boson at LEP1 or SLC. 
The excellent agreement between the current experimental results 
and the SM predictions at the quantum level implies that the 
mixing angle $\xi$ have to be small. 
In the limit of small $\xi$, the interaction Lagrangians 
for the processes 
$Z_{1,2} \to f_\alpha \ov{f_\alpha}$ are expressed as 
\begin{subequations}
\begin{eqnarray}
{\cal L}_{Z_1} &=& -\sum_{f,\, \alpha} g^{}_Z 
	\overline{f_{\alpha}} \gamma^{\mu} \left[
	\left( I^{3}_{f_{L}}-Q_{f_{\alpha}}\sin^2\theta_W 
	\right) + \tilde{Q}^{f_{\alpha}}_E \xibar \right]
	f_{\alpha}  Z_{1 \mu}, 
\label{eq:neutral1}\\ 
{\cal L}_{Z_2} &=& -\sum_{f,\,\alpha}\frac{g^{}_E}{c^{}_{\chi}}
	\overline{f_{\alpha}}\gamma^{\mu} \left[ \tilde{Q}^{f_{\alpha}}_E
	-\left( I^3_{f_{\alpha}} - Q_{f_{\alpha}}
	\sin^2\theta_W \right) \frac{g^{}_Z c^{}_{\chi}}{g^{}_E}
	\xi \right]f_{\alpha} Z_{2 \mu}, 
\label{eq:neutral2}
\end{eqnarray}
\label{eq:neutralboth}
\end{subequations}
\hsp{-0.3} 
where the effective mixing angle $\xibar$ 
in eq.~(\ref{eq:neutral1}) is given as 
\beq
\xibar = \frac{g_E}{g_Z\cos \chi } \xi. 
\eeq
In eq.~(\ref{eq:neutralboth}), the effective ${\rm U(1)'}$ 
charge $\tilde{Q}_E^{f_\alpha}$ is introduced as 
a combination of $Q_E^{f_\alpha}$ and the hypercharge $Y_{f_\alpha}$: 
\bsub
\bea
\tilde{Q}^{f_\alpha}_E &\equiv& Q^{f_\alpha}_E + Y_{f_\alpha} \delta, 
\label{eq:effective_charge}\\
\delta &\equiv&  -\frac{g^{}_Z}{g^{}_E}s^{}_W s^{}_{\chi}, 
\eea
\label{eq:u1_charge}
\esub
\hsp{-0.3}
where the hypercharge $Y_{f_\alpha}$ should be read from Table~1 
in the same manner with $Q_E^{f_\alpha}$ 
(see, eq.~(\ref{eq:charge_rule})).  
As a notable example, one can see from Table~1 
that the effective charge $\tilde{Q}_E^{f_\alpha}$ of 
the leptons ($L$ and $e^c$) disappears in the 
$\eta$-model if $\delta$ is taken to be $1/3$~\cite{eta_model}. 

Now, due to the $Z$-$Z'$ mixing, the observed 
$Z$ boson mass $m_{Z_1}$ at LEP1 or SLC is shifted from 
the SM $Z$ boson mass $m_Z$: 
\beq
\Delta m^2 \equiv m_{Z_1}^2 - m_Z^2 \le 0. 
\label{eq:mass_shift}
\eeq
The presence of the mass shift affects the 
$T$-parameter~\cite{peskin_takeuchi} at tree level. 
Following the notation of ref.~\cite{hhkm}, the $T$-parameter 
is expressed in terms of the effective form factors $\gzbar(0), 
\gwbar(0)$ and the fine structure constant $\alpha$:
\bsub
\bea
\alpha T &\equiv& 1 - \frac{\bar{g}^2_{W}(0)}{m^2_W}
	\frac{m^2_{Z_1}}{\bar{g}^2_Z(0)}  \\ 
	&=&
	\alpha \left(T_{\rm SM}^{}+  T_{\rm new}^{}\right), 
\eea
\esub
where 
$T_{\rm SM}^{}$ and the new physics contribution 
$T_{\rm new}$ are given by: 
\bsub
\bea
\alpha T_{\rm SM} 
	&=& 1 - \frac{\bar{g}^2_{W}(0)}{m^2_W}
		\frac{m^2_{Z}}{\bar{g}^2_Z(0)}, \\
\alpha T_{\rm new}
	& = & -\frac{ \Delta m^2}{m^2_{Z_1}} \geq 0. 
\eea
\esub
It is worth noting that the sign of $T_{\rm new}$ is 
always positive. 
The effects of the $Z$-$Z'$ mixing in the $Z$-pole experiments 
have hence been parametrized by the effective mixing angle 
$\xibar$ and the positive parameter $T_{\rm new}$. 

We note here that we retain 
the kinetic mixing term $\delta$ as a part of the effective $Z_1$ 
coupling $\tilde{Q}_E^{f_\alpha}$ in eq.~(\ref{eq:effective_charge}). 
As shown in refs.~\cite{eta_model,general_zzmixing,holdom2}, 
the kinetic mixing term $\delta$ can be absorbed into a further 
redefinition of $S$ and $T$. 
Such re-parametrization may be useful 
if the term $Y_{f_\alpha} \delta$ in eq.~(\ref{eq:effective_charge}) 
is much larger than the $Z'$ charge $Q_E^{f_\alpha}$. 
In the $E_6$ models studied in this paper, we find no merit in 
absorbing the $Y_f \delta$ term because, the remaining 
$Q_E^{f_\alpha}$ term is always significant. 
We therefore adopt $\tilde{Q}_E^{f_\alpha}$ as the effective 
$Z_1$ couplings and $T_{\rm new}$ accounts only for the 
mass shift (\ref{eq:mass_shift}). 
All physical consequences such as the bounds on $\xibar$ and 
$m_{Z_2}$ are of course independent of our choice of the 
parametrization. 

The two parameters $T_{\rm new}$ and $\xibar$ 
are complicated functions of the parameters of the effective 
Lagrangian (\ref{eq:l_gauge}). 
In the small mixing limit, 
we find the following useful expressions 
\bsub
\bea
\xibar &=& -\biggl( \frac{g_E}{g_Z}\frac{m_Z}{m_{Z'}} 
	\biggr)^2 \zeta \biggl[ 1+ O(\frac{m_Z^2}{m_{Z'}^2})\biggr], 
\\
\alpha T_{\rm new} &=& \hph \biggl( \frac{g_E}{g_Z}
		\frac{m_Z}{m_{Z'}} \biggr)^2 \zeta^2
	\biggl[ 1+ O(\frac{m_Z^2}{m_{Z'}^2})\biggr], 
\eea
\label{eq:tnew_xibar}
\esub
\hsp{-0.3}
where we introduced an effective mixing parameter $\zeta$ 
\beq
\zeta = \frac{g_Z}{g_E}\frac{m_{ZZ'}^2}{m_Z^2} - \delta.  
\label{eq:zeta}
\eeq
The $Z$-$Z'$ mixing effect disappears at $\zeta = 0$. 
Stringent limits on $m_{Z'}$ and hence on $m_{Z_2}$ 
can be obtained through the mixing effect if $\zeta$ is 
$O(1)$. 
We will show in Sec.~5 that $\zeta$ is calculable 
once the particle spectrum of the model is specified. 
The parameter $\zeta$ plays an essential role in 
the analysis of $Z'$ models. 

In the low-energy neutral current processes, effects of 
the exchange of the heavier mass eigenstate $Z_2$ can be 
detected. 
In the small $\xibar$ limit, they constrain the contact 
term $g_E^2/c_\chi^2 m_{Z_2}^2$. 
%
%
%
%
\section{Electroweak observables in the $Z'$ model}
\clean
In this section, we give the theoretical predictions for the 
electroweak observables which are used in our analysis. 
The experimental data of the $Z$-pole experiments and the $W$ 
boson mass measurement~\cite{lep_slc_97} 
are summarized in Table~2. 
Those for the low-energy experiments~\cite{chm} are listed in 
Table~\ref{table:low_energy}. 
\lep_table 
\lenc_table 
\subsection{Observables in $Z$-pole experiments}
The decay amplitude for the process 
$Z^{}_1 \to f_\alpha \ov{f_\alpha}$ is written as
\begin{equation}
T(Z_1 \rightarrow f_{\alpha} \ov{f_{\alpha}}) 
	= M^f_{\alpha}~\epsilon_{Z_1} \cdot J_{f_{\alpha}}, 
\label{eq:decay_amp}
\end{equation}
where $\epsilon_{Z_1}^\mu$ is the polarization vector of the 
$Z^{}_1$ boson and 
$J^{\mu}_{f_{\alpha}} = \ov{f_\alpha}\gamma^\mu f_\alpha$ is 
the fermion current without the coupling factors. 
The pseudo-observables of the $Z$-pole experiments are 
expressed in terms of the real scalar amplitudes $M_\alpha^f$ 
with the following normalization~\cite{lep_slc_97} 
\begin{equation}
g^f_{\alpha} = \frac{M^f_{\alpha}}{\sqrt{4\sqrt{2}G_Fm^2_{Z_1}}} 
	\approx \frac{M^f_{\alpha}}{0.74070}. 
\end{equation}

Following our parametrization of the $Z$-$Z'$ mixing in 
eq.~(\ref{eq:neutral1}), the effective coupling $g^f_{\alpha}$ 
in the $Z'$ models can be expressed as 
\bea
g_\alpha^f = (g_\alpha^f)_{\rm SM} + \tilde{Q}_E^{f_\alpha} \xibar. 
\eea
The SM predictions~\cite{hhkm, hhm} for the effective 
couplings $(g_\alpha^f)_{\rm SM}$ can be parametrized as 
\begin{subequations}
\begin{eqnarray}
(g^{\nu}_L)_{\rm SM} &=& \makebox[3.3mm]{ } 0.50214 
	+ 0.453\,{\Delta \bar{g}^2_Z},
\label{eq:amp_nu} \\
(g^e_L)_{\rm SM} &=& - 0.26941 - 0.244\,{\Delta \bar{g}^2_Z} 
	+ 1.001\,{\Delta \bar{s}^2},  
\label{eq:amp_el}\\
(g^e_R)_{\rm SM} &=& \makebox[3.3mm]{ } 0.23201 + 0.208\,
	{\Delta \bar{g}^2_Z} + 1.001\,{\Delta \bar{s}^2}, 
\label{eq:amp_er}\\
(g^u_L)_{\rm SM} &=&  \makebox[3.3mm]{ } 0.34694 + 0.314\,
	{\Delta \bar{g}^2_Z} - 0.668\,{\Delta \bar{s}^2}, 
\label{eq:amp_ul}\\
(g^u_R)_{\rm SM} &=& - 0.15466 - 0.139\,{ \Delta \bar{g}^2_Z}  
	- 0.668 \,{\Delta \bar{s}^2},  
\label{eq:amp_ur}\\
(g^d_L)_{\rm SM} &=& - 0.42451 - 0.383\,{\Delta \bar{g}^2_Z} 
	+ 0.334\,{\Delta \bar{s}^2}, 
\label{eq:amp_dl}\\
(g^d_R)_{\rm SM} &=& \makebox[3.3mm]{ }  0.07732
	+ 0.069\,{ \Delta \bar{g}^2_Z} + 0.334\,{\Delta \bar{s}^2},  
\label{eq:amp_dr}\\
(g^b_L)_{\rm SM} &=& - 0.42109 - 0.383\,{ \Delta \bar{g}^2_Z} 
	+ 0.334\,{\Delta \bar{s}^2} 
	+ 0.00043 x_t, 
\label{eq:amp_bl}
\end{eqnarray}
\label{eq:amp_sm}
\end{subequations}
\hsp{-0.3}
where the SM radiative corrections are expressed in terms of 
the effective couplings $\dgzbar$ and $\dsbar$~\cite{hhkm, hhm} 
and the top-quark mass dependence of the $Zb_L^{} b_L^{}$ vertex 
correction in $(g^b_L)_{\rm SM}$ is parametrized by 
the parameter $\xt$ 
\beq
\xt \equiv \frac{m_t - 175~{\rm GeV}}{10~{\rm GeV}}. 
\eeq
The gauge boson propagator corrections, $\dgzbar$ and $\dsbar$, 
are defined as the shift in the effective couplings 
$\gzbar(m_{Z_1}^2)$ and $\sbar(m_{Z_1}^2)$~\cite{hhkm} 
from their SM reference values at $\mt = 175~{\rm GeV}$ and 
$\mh = 100~{\rm GeV}$. 
They can be expressed in terms of the $S$ and $T$ parameters as 
\bsub
\begin{eqnarray}
\!\!\!\!\!\!\!\! \Delta \bar{g}^2_Z &=&  \gzbar(m_{Z_1}^2) - 0.55635 
	= 0.00412 \Delta T + 0.00005[1-(100~{\rm GeV}/m^{}_H)^2], \\  
\!\!\!\!\!\!\!\! \Delta \bar{s}^2 &=& \sbar(m_{Z_1}^2) - 0.23035 
	= 0.00360 \Delta S - 0.00241 \Delta T - 0.00023 \xa, 
\end{eqnarray}
\esub
where the expansion parameter $\xa$ is introduced to estimate 
the uncertainty of the hadronic contribution to the QED coupling 
$1/\ov{\alpha}(m_{Z_1}^2) = 128.75 \pm 0.09$~\cite{eidelman}:
\beq
\xa \equiv \frac{1/\ov{\alpha}(m_{Z_1}^2) - 128.75}{0.09}. 
\label{eq:xa_qed}
\eeq
Here, $\ds,\dt,\du$ parameters are also measured from their 
SM reference values and they are given as the sum of the SM 
and the new physics contributions 
\begin{equation}
\Delta S = \Delta S^{}_{\rm SM}+S^{}_{\rm new}, \;\;
\Delta T = \Delta T^{}_{\rm SM}+T^{}_{\rm new}, \;\; 
\Delta U = \Delta U^{}_{\rm SM}+U^{}_{\rm new}. 
\label{eq:stu_delta}
\end{equation}
The SM contributions can be parametrized as~\cite{hhm}
\begin{subequations}
\begin{eqnarray}
\Delta S^{}_{\rm SM} &=& - 0.007 x^{}_t +0.091 x^{}_H -0.010 x^2_H , \\
\Delta T^{}_{\rm SM} &=& (0.130 - 0.003 x^{}_H) x^{}_t +0.003 x^{}_t 
- 0.079 x^{}_H -0.028 x^2_H \nonumber \\ & & +0.0026 x^3_H, \\
\Delta U^{}_{\rm SM} &=& 0.022 x^{}_t -0.002x^{}_H, 
\end{eqnarray}
\end{subequations}
where $\xh$ is defined by
\begin{eqnarray}
\xh &\equiv& \log (\mh/100~{\rm GeV}). 
\label{eq:higgs_xh}
\end{eqnarray}

The pseudo-observables of the $Z$-pole experiments 
are given by using the above eight effective couplings $g_\alpha^f$ 
as follows. 
The partial width of $Z_1$ boson is given by 
\begin{eqnarray}
\Gamma_f &=& \frac{G_Fm_{Z_1}^{3}}{3\sqrt{2} \pi} \left\{ 
	\left| g^f_L + g^f_R \right|^2\frac{C_{fV}}{2} 
	+ \left| g^f_L - g^f_R \right|^2 \frac{C_{fA}}{2}  \right\}
	\left( 1+\frac{3}{4}Q^2_f\frac{\bar{\alpha}(m^2_{Z_1})}{\pi}\right)
        \makebox[10mm][l]{,}
\label{eq:partial_width}
\end{eqnarray}
where the factors $C_{fV}$ and $C_{fA}$ account for the 
finite mass corrections and the final state QCD corrections 
for quarks. 
Their numerical values are listed in Table~\ref{tab:cvca}. 
The $\alpha_s$-dependence in $C_{qV}, C_{qA}$ 
is parametrized in terms of the parameter $\xs$
\beq
\xs \equiv \frac{\alpha_s(m_{Z_1})-0.118}{0.003}. 
\eeq
The last term proportional to $\abar(m^2_{Z_1})/\pi$ 
in eq.~(\ref{eq:partial_width}) accounts for the 
final state QED correction. 
\cvca_tab
The total decay width $\Gamma_{Z_1}$ and the hadronic decay 
width $\Gamma_h$ are given in terms of $\Gamma_f$: 
\bsub
\begin{eqnarray}
\Gamma_{Z_1} &=& 3\Gamma_{\nu} + \Gamma_e 
	+ \Gamma_{\mu} + \Gamma_{\tau} + \Gamma_h, 
\label{eq:total_width}\\
\Gamma_h &=& \Gamma_u + \Gamma_c + \Gamma_d + \Gamma_s + \Gamma_b. 
\label{eq:hadron_width}
\end{eqnarray}
\esub
The ratios $R_\ell^{}, R_c^{}, R_b^{}$ and the hadronic peak 
cross section $\sigma_h^0$ are given by: 
\begin{equation}
R_{\ell} = \frac{\Gamma_h}{\Gamma_e},\;
R_c      = \frac{\Gamma_c}{\Gamma_h},\;
R_b      = \frac{\Gamma_b}{\Gamma_h},\;
\sigma^0_h = \frac{12\pi}{m^2_{Z_1}}
	\frac{\Gamma_e\Gamma_h}{\Gamma_{Z_1}^2}. 
\end{equation}

The left-right asymmetry parameter $A^{f}$ is also 
given in terms of the effective couplings $g_\alpha^f$ as  
\begin{equation}
A^f 	= \frac{(g^{f}_L)^2-(g^{f}_R)^2}{(g^{f}_L)^2+(g^{f}_R)^2}. 
\end{equation}
The forward-backward (FB) asymmetry $A^{0,f}_{FB}$ 
and the left-right (LR) asymmetry $A^{0,f}_{LR}$ 
are then given as follows: 
\bsub
\bea
A^{0,f}_{FB} &=& \frac{3}{4}A^{e}A^{f}, \\
A^{0,f}_{LR} &=&  A^{f}. 
\eea
\esub

\subsection{$W$ boson mass}
The theoretical prediction of $m^{}_W$
can be parametrized as~\cite{hhkm,hhm}
\begin{equation}
m_W^{}{\rm (GeV)} =80.402-0.288\,{\it \Delta S}+0.418\,{\it \Delta T} 
+0.337\,{\it \Delta U}+0.012\,{\it x_{\alpha}}, 
\end{equation}
by using the same parameters, $\ds, \dt, \du$ (\ref{eq:stu_delta}) 
and $\xa$ (\ref{eq:xa_qed}). 

\subsection{Observables in low-energy experiments} 
\clean
In this subsection, we show the theoretical predictions for the 
electroweak observables in the low-energy neutral current 
experiments (LENC) --- 
(i) polarization asymmetry of the charged lepton scattering off 
nucleus target (\ref{section_slac}--\ref{section_mainz}), 
(ii) parity violation in cesium atom (\ref{section_apv}),  
(iii) inelastic $\nu_\mu$-scattering off nucleus target 
(\ref{section_nq}) and 
(iv) neutrino-electron scattering (\ref{section_ne}). 
The experimental data are summarized in Table~\ref{table:low_energy}. 
Theoretical expressions for the observables of 
(i) and (ii) are conveniently given in terms of 
the model-independent parameters $C_{1q}, C_{2q}$~\cite{jekim} 
and $C_{3q}$~\cite{chm}.  
The $\nu_\mu$-scattering data (iii) and (iv) are expressed 
in terms of the parameters $g_{L\alpha}^{\nu_\mu f}$. 
All the model-independent parameters can be expressed 
compactly in terms of the reduced helicity amplitudes 
$M_{\alpha\beta}^{f f'}$~\cite{chm,hhkm} of the 
process $f_\alpha f'_\beta \to f_\alpha f'_\beta$: 
\bsub
\bea
C_{1q} &=& \frac{1}{2 \sqrt{2} G_F} ( \hph M_{LL}^{\ell q}
	+ M_{LR}^{\ell q} - M_{RL}^{\ell q} - M_{RR}^{\ell q} ), 
\\
C_{2q} &=& \frac{1}{2 \sqrt{2} G_F} ( \hph M_{LL}^{\ell q}
	- M_{LR}^{\ell q} + M_{RL}^{\ell q} - M_{RR}^{\ell q} ), 
\\
C_{3q} &=& \frac{1}{2 \sqrt{2} G_F} ( 	-M_{LL}^{\ell q}
	+ M_{LR}^{\ell q} + M_{RL}^{\ell q} - M_{RR}^{\ell q} ), 
\\
g_{L\alpha}^{\nu_\mu f} &=& \frac{1}{2 \sqrt{2} G_F} 
	(-M_{L\alpha}^{\nu_\mu f} ). 
\label{eq:nutrino_amplitude}
\eea
\esub
Below, we divide these model-independent parameters into two pieces as 
\bsub
\bea
C_{iq} &=& (C_{iq})_{\rm SM} + \Delta C_{iq}, \\
g_{L\alpha}^{\nu_\mu f} &=& (g_{L\alpha}^{\nu_\mu f})_{\rm SM} + 
	\Delta g_{L\alpha}^{\nu_\mu f}, 
\eea
\esub
where the first term in each equation is the SM contribution
which is parametrized conveniently by $\ds$ and $\dt$ in 
ref.~\cite{chm}. 
The terms $\Delta C_{iq}$ and $\Delta g_{L\alpha}^{\nu_\mu f}$ 
represent the additional contributions from the $Z$-$Z'$ mixing 
and the $Z_2$ exchange:
\begin{subequations}
\begin{eqnarray}
\Delta C^{}_{1u} &=& 
	(-0.19s^{}_\beta-0.15c^{}_\beta+0.65\delta )\bar{\xi} 
	-\frac{g^2_E}{c^2_\chi}
	\frac{(\tilde{Q}^L_E-\tilde{Q}^E_E)
	(\tilde{Q}^Q_E+\tilde{Q}^U_E)}	{2\sqrt{2}G^{}_Fm^2_{Z_2}}, \\	
\Delta C^{}_{1d} &=& 
	(0.36s^{}_\beta-0.54c^{}_\beta+0.17\delta )\bar{\xi}
	-\frac{g^2_E}{c^2_\chi}
	\frac{(\tilde{Q}^L_E-\tilde{Q}^E_E)(\tilde{Q}^Q_E+\tilde{Q}^D_E)}
	{2\sqrt{2}G^{}_Fm^2_{Z_2}}, \\
\Delta C^{}_{2u} &=& 
	(0.02s^{}_\beta-0.84c^{}_\beta+1.48\delta )\bar{\xi} 
	-\frac{g^2_E}{c^2_\chi}\frac{(\tilde{Q}^L_E+\tilde{Q}^E_E)
	(\tilde{Q}^Q_E-\tilde{Q}^U_E)}{2\sqrt{2}G^{}_Fm^2_{Z_2}}, \\
\Delta C^{}_{2d} &=&
	(0.02s^{}_\beta+0.84c^{}_\beta-1.48\delta )\bar{\xi}
	-\frac{g^2_E}{c^2_\chi}\frac{(\tilde{Q}^L_E+\tilde{Q}^E_E)
	(\tilde{Q}^Q_E-\tilde{Q}^D_E)}{2\sqrt{2}G^{}_Fm^2_{Z_2}}, \\
\Delta C^{}_{3u} &=&
	(-0.82c^{}_\beta+1.00\delta )\bar{\xi}
	-\frac{g^2_E}{c^2_\chi}\frac{(\tilde{Q}^L_E-\tilde{Q}^E_E)
	(\tilde{Q}^U_E-\tilde{Q}^Q_E)}{2\sqrt{2}G^{}_Fm^2_{Z_2}}, \\
\Delta C^{}_{3d} &=&
	(1.06s^{}_\beta-0.82c^{}_\beta-1.00\delta )\bar{\xi}
	-\frac{g^2_E}{c^2_\chi}\frac{(\tilde{Q}^L_E-\tilde{Q}^E_E)
	(\tilde{Q}^D_E-\tilde{Q}^Q_E)}{2\sqrt{2}G^{}_Fm^2_{Z_2}}, \\
\Delta g_{LL}^{\nu u} &=& 
	(0.44s^{}_\beta+0.22c^{}_\beta-0.18\delta )\bar{\xi}
	+\frac{g^2_E}{c^2_\chi}\frac{\tilde{Q}^L_E \tilde{Q}^Q_E}
	{2\sqrt{2}G^{}_Fm^2_{Z_2}}, \\
\Delta g_{LR}^{\nu u} &=& 
	(-0.35s^{}_\beta+0.01c^{}_\beta+0.82\delta )\bar{\xi}
	+\frac{g^2_E}{c^2_\chi}\frac{\tilde{Q}^L_E \tilde{Q}^U_E}
	{2\sqrt{2}G^{}_Fm^2_{Z_2}}, \\
\Delta g_{LL}^{\nu d} &=& 
	(0.04s^{}_\beta-0.72c^{}_\beta+0.59\delta )\bar{\xi}
	+\frac{g^2_E}{c^2_\chi}\frac{\tilde{Q}^L_E \tilde{Q}^Q_E}
	{2\sqrt{2}G^{}_Fm^2_{Z_2}}, \\
\Delta g_{LR}^{\nu d} &=& 
	(-0.22s^{}_\beta-0.52c^{}_\beta-0.41\delta )\bar{\xi}
	+\frac{g^2_E}{c^2_\chi}\frac{\tilde{Q}^L_E \tilde{Q}^D_E}
	{2\sqrt{2}G^{}_Fm^2_{Z_2}},\\
%
\Delta g_{LL}^{\nu e} &=& (0.12 s^{}_\beta + 0.28 c^{}_\beta 
	- 0.23 \delta) \xibar 
	+\frac{g^2_E}{c^2_\chi}\frac{\tilde{Q}^L_E \tilde{Q}^L_E}
	{2\sqrt{2}G^{}_Fm^2_{Z_2}},
\\
\Delta g_{LR}^{\nu e} &=& (-0.14 s^{}_\beta + 0.49 c^{}_\beta 
	- 1.23 \delta) \xibar
	+\frac{g^2_E}{c^2_\chi}\frac{\tilde{Q}^L_E \tilde{Q}^e_E}
	{2\sqrt{2}G^{}_Fm^2_{Z_2}}. 
\end{eqnarray}
\label{eq:lenc_extra}
\end{subequations}
where $c^{}_\beta = \cos\beta_E$ and $s^{}_\beta = \sin\beta_E$. 

\subsubsection{SLAC $e$D experiment}
\label{section_slac}
The parity asymmetry in the inelastic scattering of polarized 
electrons from the deuterium target was measured at SLAC~\cite{slac}. 
The experiment constrains the parameters 
$2C_{1u}-C_{1d}$ and $2C_{2u}-C_{2d}$. 
The most stringent constraint shown in Table~\ref{table:low_energy} 
is found for the following combination
\begin{subequations}
\begin{eqnarray}
A_{\rm SLAC} &=& 2C_{1u}-C_{1d} +0.206(2C_{2u}-C_{2d}) \\
	&=& 0.745 - 0.016\,{\it \Delta S} + 0.016\,{\it \Delta T} 
	\nonumber \\
	&&~~~~
	+ 2\Delta C_{1u}- \Delta C_{1d} 
	+ 0.206(2\Delta C_{2u}- \Delta C_{2d}), 
\end{eqnarray}
\label{eq:slac_sm1}
\end{subequations}
where the theoretical prediction~\cite{chm} is evaluated 
at the mean momentum transfer $\langle Q^2 \rangle = $ 1.5 GeV$^2$. 

\subsubsection{CERN $\mu^\pm$C experiment}
The CERN $\mu^\pm$C experiment~\cite{cern} measured 
the charge and polarization asymmetry of deep-inelastic 
muon scattering off the ${}^{12}$C target.
The mean momentum transfer of the experiment may be estimated 
at $\langle Q^2 \rangle = $ 50 GeV$^2$~\cite{souder}.
The experiment constrains the parameters 
$2C_{2u}-C_{2d}$ and $2C_{3u}-C_{3d}$. 
The most stringent constraint is found for the following 
combination~\cite{chm}
\begin{subequations}
\begin{eqnarray}
A_{\rm CERN} &=& 2C_{3u}-C_{3d}+0.777(2C_{2u}-C_{2d}) \\
	 &=& -1.42-0.016\,{\it \Delta S}+0.0006\,{\it \Delta T}  
	\nonumber \\
	&&~~~+ 2\Delta C_{3u}- \Delta C_{3d}
	+ 0.777(2\Delta C_{2u}- \Delta C_{2d}). 
\end{eqnarray}
\label{eq:cern_sm1}
\end{subequations}
\subsubsection{Bates $e$C experiment} 
The polarization asymmetry of the electron elastic scattering
off the ${}^{12}$C target was measured at Bates \cite{bates}.
The experiment constrains the combination
\bsub
\bea
A_{\rm Bates} &=& C_{1u}+C_{1d} \\
	&=& - 0.1520 - 0.0023\,{\it \Delta S} 
	+ 0.0004\,{\it \Delta T} 
	+ \Delta C_{1u} + \Delta C_{1d}, 
\eea
\esub
where the theoretical prediction~\cite{chm} is evaluated 
at $\langle Q^2 \rangle = $ 0.0225 GeV$^2$.
\subsubsection{Mainz $e$Be experiment }
\label{section_mainz}
The polarization asymmetry of electron quasi-elastic scattering
off the ${}^9$Be target was measured at Mainz \cite{mainz}.
The data shown in Table~\ref{table:low_energy} is for 
the combination
\bsub
\bea
A_{\rm Mainz} &=& -2.73 C_{1u} + 0.65 C_{1d} - 2.19 C_{2u} 
	+ 2.03 C_{2d} 
\\
	&=&  -0.876 + 0.043\Delta S - 0.035\Delta T
	\nonumber \\
	&&~~~
	-2.73 \Delta C_{1u} + 0.65 \Delta C_{1d} 
	- 2.19 \Delta C_{2u} + 2.03 \Delta C_{2d}, 
\eea
\esub
where the theoretical prediction~\cite{chm} is evaluated 
at $\langle Q^2 \rangle = $ 0.2025 GeV$^2$. 
\subsubsection{Atomic Parity Violation}
\label{section_apv}
The experimental results of parity violation in the atom 
are often given in terms of the weak charge $Q^{}_W(A,Z)$
of nuclei. By using the model-independent parameter $C_{1q}^{}$,
the weak charge of a nuclei can be expressed as 
\begin{equation}
Q^{}_W(A,Z)=2ZC_{1p}^{}+2(A-Z)C_{1n}^{}. 
\end{equation}
By taking account of the long-distance photonic 
correction~\cite{apv_photonic}, 
we find $C_{1p}$ and $C_{1n}$ as 
\begin{subequations}
\begin{eqnarray}
C_{1p} &=& \hph 0.03601  -0.00681 \Delta S + 0.00477\,\Delta T 
	+ 2\Delta C_{1u} + \Delta C_{1d}, \\
C_{1n} &=& -0.49376 - 0.00366\,{\it \Delta T} 
	+ \Delta C_{1u} + 2\Delta C_{1d}. 
\end{eqnarray}
\end{subequations}
The data for cesium atom $^{133}_{55}Cs$~\cite{noecker,wood} 
is given in Table~\ref{table:low_energy} and 
the theoretical prediction of the weak charge 
is found to be~\cite{chm} 
\begin{equation}
Q_{W}(^{133}_{55}Cs) =  -73.07 -0.749\,{\it \Delta S}
	- 0.046\,{\it \Delta T} 
	+ 376 \Delta C_{1u} + 422 \Delta C_{1d}. 
\end{equation}
\subsubsection{Neutrino-quark scattering} 
\label{section_nq}
For the $\nu_\mu$-quark scattering, the experimental results 
up to the year 1988 were summarized in ref.~\cite{fh} in terms 
of the model-independent parameters $g_L^2, g_R^2, \delta_L^2, \delta_R^2$. 
The most stringent constraint on the result in ref.~\cite{fh} 
is found for the 
following combination: 
\bea
K_{\rm FH} &=& g_L^2 + 0.879 g_R^2 -0.010 \delta_L^2 -0.043 \delta_R^2.
\eea
More recent CCFR experiment at Tevatron measured the following 
combination~\cite{ccfr}
\begin{eqnarray}
K_{\rm CCFR} &=&  1.7897 g_L^2 + 1.1479 g_R^2 - 0.0916 \delta_L^2 
	- 0.0782 \delta_R^2. 
\end{eqnarray}
The data are shown in Table~\ref{table:low_energy} and the 
SM predictions are calculated from our reduced 
amplitudes (\ref{eq:nutrino_amplitude}) as 
follows~\cite{chm,hhkm}
\bea
g_\alpha^2 = (g_{L\alpha}^{\nu_\mu u})^2 
	+ (g_{L\alpha}^{\nu_\mu d})^2, ~~~ 
\delta_\alpha^2 = (g_{L\alpha}^{\nu_\mu u})^2 
	- (g_{L\alpha}^{\nu_\mu d})^2,  
\eea
for $\alpha = L$ and $R$, respectively, where 
\bsub
\bea
g_{LL}^{\nu_\mu u} &=& 
	\hph 0.3468 - 0.0023 \ds + 0.0041 \dt, 
\\
g_{LR}^{\nu_\mu u} &=& 
	-0.1549 - 0.0023 \ds + 0.0004 \dt, 
\\
g_{LL}^{\nu_\mu d} &=& 
	-0.4299 + 0.0012 \ds - 0.0039 \dt, 
\\
g_{LR}^{\nu_\mu d} &=& 
	\hph 0.0775 + 0.0012 \ds - 0.0002 \dt.
\eea
\esub
The above predictions are obtained at the momentum transfer 
$\langle Q^2 \rangle = 35~{\rm GeV}^2$ relevant for the 
CCFR experiment~\cite{ccfr}. 
The estimations are found to be valid~\cite{chm} also 
for the data of ref.~\cite{fh}, whose typical scale is 
$\langle Q^2 \rangle = 20~{\rm GeV}^2$. 

\subsubsection{Neutrino-electron scattering} 
\label{section_ne}
The $\nu_\mu$-$e$ scattering experiments measure the neutral 
currents in a purely leptonic channel. 
The combined results~\cite{chm,charm-II} are given in 
Table~\ref{table:low_energy}. 
The theoretical predictions 
\bsub
\bea
g_{LL}^{\nu_\mu e} &=& -0.273 + 0.0033 \ds - 0.0042 \dt 
	+ \Delta g_{LL}^{\nu_\mu e}, 
\\
g_{LR}^{\nu_\mu e} &=& \hph 0.233 + 0.0033 \ds - 0.0006 \dt 
	+ \Delta g_{LR}^{\nu_\mu e}, 
\eea
\esub
are evaluated at $\langle Q^2 \rangle = 2m_e E_\nu$ 
with $E_\nu = 25.7~{\rm GeV}$ for the CHARM-II 
experiment~\cite{charm-II}. 
%
%
%
%
\section{Constraints on $Z'$ bosons from electroweak experiments}
\clean 
Following the parametrization presented in Sec.~3, 
we can immediately obtain the constraints on $T_{\rm new}, \xibar$ 
and $\contact$ from the data listed in Table~2 
and Table~\ref{table:low_energy}. 
Setting $S_{\rm new} = U_{\rm new} = 0$, we find that 
the $Z$-pole measurements constrains $T_{\rm new}$ and $\xibar$ 
while $m_W$ data constrains $T_{\rm new}$. 
The contact term $\contact$ is constrained from the LENC data. 
The number of the free parameters is, therefore, six: 
the above three parameters and the SM parameters, 
$\mt, \alpha_s(m_{Z_1})$ and $\abar(m_{Z_1}^2)$. 
Throughout our analysis, we use 
\bsub
\bea
\mt &=& 175.6 \pm 5.5~{\rm GeV}~\cite{mt96}, \\
\alpha_s (m_{Z_1}^{}) &=& 0.118 \pm 0.003~\cite{PDG}, \\
1/\abar(m_{Z_1}^2) &=& 128.75 \pm 0.09~\cite{eidelman}, 
\eea
\esub
as constraints on the SM parameters. 
The Higgs mass dependence of the results are parametrized 
by $\xh$ (\ref{eq:higgs_xh}) 
in the range $77~{\rm GeV} <  \mh \lsim 150~{\rm GeV}$. 
The lower bound is obtained at the LEP experiment~\cite{higgs_direct}. 
The upper bound is the theoretical limit on the lightest Higgs 
boson mass in any supersymmetric models that accommodate 
perturbative unification of the gauge couplings~\cite{kane}. 
We first obtain the constraints from the $Z$-pole experiments 
and $W$ boson mass measurement only, and then obtain 
those by including the LENC experiments.  

\subsection{Constraints from $Z$-pole and $m_W$ data}
Let us examine first the constraints from the $Z$-pole and 
$m_W$ data by performing the five-parameter fit for $T_{\rm new}, 
\xibar, \mt, \alpha_s(m_{Z_1})$ and $\abar(m_{Z_1}^2)$. 
The results for the $\chi, \psi, \eta$ and $\nu$ models at 
$\delta = 0$ are summarized as follows: 
\def\labelenumi{(\theenumi)}
\def\theenumi{\roman{enumi}}
\begin{enumerate}
\item $\chi$-model ($\delta = 0$)
\bea
\begin{array}{l}
	\left.
	\begin{array}{lcl}
	T_{\rm new} &=& -0.040 + 0.15\xh \pm 0.12 \\
	\xibar &=& \hph 0.00017 - 0.00005\xh \pm 0.00046
	\end{array}
	\right \} \rho_{\rm corr} = 0.28,  \\
\chi^2_{\rm min}/({\rm d.o.f.}) = (16.5 + 0.7 \xh)/(12), 
\end{array}
\label{eq:const_chi}
\eea
\item $\psi$-model ($\delta = 0$)
\bea
\begin{array}{l}
	\left.
	\begin{array}{lcl}
	T_{\rm new} &=& -0.043 + 0.16\xh \pm 0.11 \\
	\xibar &=& \hph 0.00019 + 0.00012\xh \pm 0.00050
	\end{array}
	\right \} \rho_{\rm corr} = 0.20, \\
\chi^2_{\rm min}/({\rm d.o.f.}) = (16.5 + 0.4 \xh)/(12), 
\end{array}
\label{eq:const_psi}
\eea
\item $\eta$-model ($\delta = 0$)
\bea
\begin{array}{l}
	\left.
	\begin{array}{lcl}
	T_{\rm new} &=& -0.053 + 0.14\xh \pm 0.11 \\
	\xibar &=& -0.00014 - 0.00062\xh \pm 0.00108 
	\end{array}
	\right \} \rho_{\rm corr} = 0.09, \\
\chi^2_{\rm min}/({\rm d.o.f.}) = (16.6 + 0.4 \xh)/(12), 
\end{array}
\label{eq:const_eta}
\eea
\item $\nu$-model ($\delta = 0$)
\bea
\begin{array}{l}
	\left.
	\begin{array}{lcl}
	T_{\rm new} &=& -0.042 + 0.15\xh \pm 0.11 \\
	\xibar &=& \hph 0.00016 +  0.00007\xh \pm 0.00042 
	\end{array}
	\right \} \rho_{\rm corr} = 0.23, \\
\chi^2_{\rm min}/({\rm d.o.f.}) = (16.5 + 0.5 \xh)/(12). 
\end{array}
\label{eq:const_nu}
\eea
\end{enumerate}
In the above four $Z'$ models, the results for $T_{\rm new}$ 
and $\xibar$ are consistent with zero for $\xh = 0$. 
Moreover, the best fits of $T_{\rm new}$ in all the $Z'$ models 
are in the unphysical 
region, $T_{\rm new} < 0$. 
The parameter $T_{\rm new}$ could be positive for the large $\xh$: 
For example, $\xh = 0.41$ ($\mh = 150~{\rm GeV}$) makes $T_{\rm new}$ 
in all the four $Z'$ models positive. 
The allowed range of the effective mixing angle $\xibar$ is 
order of $10^{-3}$ for the $\eta$-model and $10^{-4}$ for 
the other three models in 1-$\sigma$ level. 
The $\xh$-dependence of $\xibar$ in the $\eta$-model is 
larger than the other three models. 
For comparison, we show the result for the leptophobic 
$\eta$-model ($\delta=1/3$) 
\begin{enumerate}
\addtocounter{enumi}{4}
\item leptophobic $\eta$-model ($\delta = 1/3$)
\bea
\begin{array}{l}
	\left.
	\begin{array}{lcl}
	T_{\rm new} &=& -0.049 + 0.15\xh \pm 0.11 \\
	\xibar &=& \hph 0.00269 +  0.00026\xh \pm 0.00309 
	\end{array}
	\right \} \rho_{\rm corr} = 0.03, \\
\chi^2_{\rm min}/({\rm d.o.f.}) = (15.9 + 0.5 \xh)/(12).
\end{array}
\label{eq:zpole_leptophobic}
\eea
\end{enumerate}
By comparing the $\eta$-model with no kinetic mixing ($\delta = 0$) 
in eq.~(\ref{eq:const_eta}), we find significantly 
weaker constraint on $\xibar$. 
\tnew_xi 
In Fig.~\ref{allowed_eta}, 
we show the 1-$\sigma$ and 90\% CL allowed region on 
the $(\xibar, T_{\rm new})$ plane 
in the $\eta$-model with $\delta = 0$ and $1/3$ for 
$\mh = 100~{\rm GeV}$. 

The best fit results at $\mh = 100~{\rm GeV}$ under the 
constraint $T_{\rm new} \geq 0$ are shown in 
Table~2. 
We can see from Table~2 that 
there is no noticeable improvement of the fit 
for the $\chi,\psi,\eta$ and $\nu$ models at $\delta = 0$. 
The $\chi^2_{\rm min}$ remains almost the same as that of the SM, 
even though each model has two new free parameters, 
$T_{\rm new}$ and $\xibar$. 
The fit slightly improves for the leptophobic $\eta$-model 
($\delta = 1/3$) because of the smaller pull factor 
for the $R_b$ data. 
The probability of the fit, 18.7\% CL, is still less 
than that of the SM, 26.2\% CL, because the $\chi^2_{\rm min}$ 
reduces only 0.8 despite two additional free parameters. 

We explore the whole range of the parameters, $\beta_E$ and $\delta$. 
In Fig.~\ref{chisq_distribution}, we show the improvement in 
$\chi^2_{\rm min}$ over the SM value, $\chi^2_{\rm min}({\rm SM})
= 16.9$ (see Table~2): 
\bea
\Delta \chi^2 \equiv \chi_{\rm min}^2(\beta_E, \delta) 
	- \chi^2_{\rm min}({\rm SM}), 
\eea
where $\chi^2_{\rm min}(\beta_E, \delta)$ is evaluated 
at the specific value of $\beta_E$ and $\delta$ for 
$\mh = 100~{\rm GeV}$.
\fig_beta_delta 
As we seen from Fig.~\ref{chisq_distribution}, the $\chi^2_{\rm min}$ 
depends very mildly in the whole range of the $\beta_E$ and $\delta$ 
plane, except near the leptophobic $\eta$-model 
($\beta_E = \tan^{-1}(\sqrt{5/3})$ and $\delta = 1/3$)~\cite{eta_model}. 
Even for the best choice of $\beta_E$ and $\delta$, the improvement in 
$\chi^2_{\rm min}$ is only 1.5 over the SM. 
Because each model has two additional parameters $T_{\rm new}$ and 
$\xibar$, we can conclude that no $Z'$ model in this framework 
improves the fit over the SM. 
The ``$\times$'' marks plotted in Fig.~\ref{chisq_distribution} show 
the specific models which we will discuss in the next section. 

\subsection{Constraints from $Z$-pole + $m_W$ + LENC data}
Next we find constraints on the contact term $\contact$ 
by including the low-energy data in addition to the 
$Z$-pole and $m_W$ data.
Because $T_{\rm new}$ and $\xibar$ are already constrained 
severely by the $Z$-pole and $m_W$ data, only the contact 
terms proportional to $g_E^2/c_\chi^2 m_{Z_2}^2$ 
contribute to 
the low-energy observables, except for the special case 
of the leptophobic $\eta$-model ($\delta = 1/3$). 

We summarize the results of the six-parameter 
fit for the $\psi, \chi, \eta$ and $\nu$ models:
\begin{flushleft}
(i) $\chi$-model
\bsub
\bea
& & \!\!\!\!\left. \!\!
\begin{array}{r c l c l c l }
T_{\rm new} &\!\!=\!\!&\!\! -0.063\!\!&\!\!+\!\!&\!\! 0.14  x^{}_H\!\!
&\!\!\pm\!\!&\!\!0.11\!\!\!\!  \\
\bar{\xi}   &\!\!=\!\!&\!\! -0.00005\!\!&\!\!-\!\!&\!\!0.00006x^{}_H\!\!
&\!\!\pm\!\!&\!\! 0.00044\!\!\! \\
g^2_E/c^2_{\chi}m^2_{Z_2} &\!\!=\!\!& 
\makebox[3.3mm]{}\!\!0.26\!\!&\!\!+\!\!&
\!\!0.01 x^{}_H\!\! &\!\!\pm\!\!&\!\!0.21\!\!
\end{array} \right\} 
\rho_{\rm corr} \!=\! \left(
\begin{array}{rrr}
\!\!1.00\!&\!0.25\!& \!0.09\!\! \\
        &\!1.00\!& \!0.15\! \\
        &        & \!1.00\!
\end{array} \right)\!\!,\makebox[6mm]{ }  
 \\ \!\!\!\!\!&&\chi^2_{\rm min}/ ({\rm d.o.f.}) 
= (19.9 + 0.9 x^{}_H)/(20), 
\eea
\esub
\end{flushleft}
\begin{flushleft}
(ii) $\psi$-model
\bsub
\bea
& & \!\!\!\!\left. \!\!
\begin{array}{r c l c l c l }
T_{\rm new} &\!\!=\!\!&\!\! -0.065 \!\!  &
\!\!+\!\!&\!\! 0.15  x^{}_H\!\!&\!\!\pm\!\!&\!\!0.11\!\! \\
\bar{\xi}   &\!\!=\!\!&\!\! -0.00014\!\! &
\!\!+\!\!&\!\! 0.00012x^{}_H\!\!&\!\!\pm\!\!&\!\!0.00050\!\!\! \\
g^2_E/c^2_{\chi}m^2_{Z_2} &\!\!=\!\!&\!\! \makebox[3.3mm]{}1.66\!\!&
\!\!+\!\!&\!\!0.19 x^{}_H\!\!&\!\!\pm\!\!&\!\! 2.90\!\!
\end{array} \right\} 
\rho^{}_{\rm corr} \!=\! \left( 
\begin{array}{rrr}
\!\!1.00\! & \!0.19\! & \!0.07\!\! \\
         & \!1.00\! & \!0.03\! \\
         &          & \!1.00\! 
\end{array} \right)\!\!,\makebox[6mm]{ }  
 \\ \!\!\!\!\!&&\chi^2_{\rm min}/ ({\rm d.o.f.}) 
	= (21.1+0.8 x^{}_H)/(20), 
\eea
\esub
\end{flushleft}
\begin{flushleft}
(iii) $\eta$-model
\bsub
\bea
\hsp{-2.5}& & \!\!\!\!\left. \!\!
\begin{array}{r c l c l c l }
T_{\rm new}\! &\!\!\!=\!\!&\!\! -0.074\!\! &\!\!+\!\!&
\!\! 0.14  x^{}_H\!\!&\!\!\pm\!\!&\!\!0.11\!\!\! \\
\bar{\xi}\!   &\!\!\!=\!\!&\!\! -0.00038\!\! &\!\!-\!\!& 
\!\!0.00063x^{}_H\!\!&\!\!\pm\!\!&\!\!0.00106\!\!\!\! \\
g^2_E/c^2_{\chi}m^2_{Z_2}\! &\!\!=\!\!&\!\!-0.62\!\!&\!\!+\!\!&
\!\!0.08 x^{}_H\!\!&\!\!\pm\!\!&\!\!0.87\!\!\!
\end{array} \right\}\! 
\rho^{}_{\rm corr} \!=\! \left( 
\begin{array}{rrr}
\!\!1.00\! & \!0.06\! & \!-0.05\!\! \\
         & \!\!1.00\! & \!-0.22\!\! \\
         &          & \!1.00\!  
\end{array} \right)\!\!,\makebox[6mm]{ }  
 \\ \!\!\!\!\!&&\chi^2_{\rm min}/ ({\rm d.o.f.}) 
	= (20.8+0.5 x^{}_H)/(20), 
\eea
\esub
\end{flushleft}
\begin{flushleft}
(iv) $\nu$-model
\bsub
\bea
\hsp{-1.5}& & \!\!\!\!\left. \!\!
\begin{array}{r c l c l c l }
T_{\rm new} &\!\!=\!\!&\!\! -0.061\!\! &\!\!+\!\!&
\!\! 0.15  x^{}_H\!\!&\!\!\pm\!\!&\!\!0.11\!\! \\
\bar{\xi}   &\!\!=\!\!&\!\! \makebox[3.3mm]{} 0.00010\!\! &\!\!+\!\!&
\!\! 0.00006x^{}_H\!\!&\!\!\pm\!\!&\!\!0.00041\!\!\!\! \\
g^2_E/c^2_{\chi}m^2_{Z_2}\! &\!\!=\!\!&\!\!-0.65\!\!&\!\!+\!\!&
\!\!0.04 x^{}_H\!\!&\!\!\pm\!\!&\!\! 0.54\!\!
\end{array} \right\} 
\rho^{}_{\rm corr} \!=\! \left( 
\begin{array}{rrr}
\!\!1.00\! & \!0.21\! & \!0.07\! \\
         & \!1.00\! & \!0.03\! \\
         &          & \!1.00\!
\end{array} \right)\!\!,\makebox[6mm]{ }  
 \\ \!\!\!\!\!&&\chi^2_{\rm min}/ ({\rm d.o.f.}) 
	= (20.1+0.8 x^{}_H)/(20).
\eea
\esub
\end{flushleft}
The contact term $\contact$ in the $\psi$ and $\eta$ models 
is consistent with zero in 1-$\sigma$ level. 
Both the best fit and the 1-$\sigma$ error of 
the parameters $T_{\rm new}$ and $\xibar$ in all the 
$Z'$ models are slightly affected 
by including the LENC data: The best fit value of $T_{\rm new}$ 
in all the $Z'$ models cannot be positive 
even for the $\mh = 150~{\rm GeV}$ $(\xh = 0.41)$. 
Since the leptophobic $\eta$-model does not have the contact term, 
the low-energy data constrain the same parameters $T_{\rm new}$ 
and $\xibar$. 
After taking into account both the high-energy and low-energy 
data, we find 
\begin{enumerate}
\addtocounter{enumi}{4}
\item leptophobic $\eta$ model ($\delta = 1/3$)
\bea
\begin{array}{l}
\left.
\begin{array}{lrl}
T_{\rm new} &=&  -0.074 + 0.148 \xh \pm 0.110\\ 
\xibar      &=&  \hph 0.00157 + 0.00019 \xh \pm 0.00279
\end{array}
	\right \}\rho_{\rm corr} = 0.02, \\
\chi^2_{\rm min}/({\rm d.o.f.}) 
	= (21.2 + 1.0 \xh)/(21). 
\end{array}
\eea
\end{enumerate}
The allowed range of $\xibar$ is slightly severe as 
compared to eq.~(\ref{eq:zpole_leptophobic}). 

The best fit results for $\mh = 100~{\rm GeV}$ under the 
condition $T_{\rm new} \geq 0$ are shown in Table~\ref{table:low_energy}. 
It is noticed that the best fit values for the weak charge of 
cesium atom $^{133}_{55}Cs$ in the $\chi, \eta$ and $\nu$ models 
are quite close to the experimental data. 
These models lead to $\Delta \chi^2 = -1.8$ ($\chi$), 
$-0.8$ $(\eta)$ and $-1.6$ $(\nu)$. 
No other noticeable point is found in the table. 
\mass_lenc 

The above constraints on $\contact$ from the LENC data give 
the lower mass bound of the heavier mass eigenstate $Z_2$ 
in the $Z'$ models except for the leptophobic $\eta$-model. 
In Fig.~\ref{mass_distribution}, the contour plot of 
the 95\% CL lower mass limit of $Z_2$ boson from the 
LENC experiments 
are shown on the $(\beta_E, \delta)$ plane by setting 
$g_E = g_Y$ and $\mh = 100~{\rm GeV}$ 
under the condition $m_{Z_2} \geq 0$. 
In practice, we obtain the 95\% CL lower limit of 
the $Z_2$ boson mass $m_{95}$ in the following way: 
\bea
0.05 =
 \frac{
\int^\infty_{m_{95}} dm_{Z_2} P(m_{Z_2})}
{\int^\infty_{0} dm_{Z_2} P(m_{Z_2})}, 
\eea
where we assume that the probability density function 
$P(m_{Z_2})$ is proportional to ${\rm exp}(-\chi^2(m_{Z_2})/2)$. 

We can read off from Fig.~\ref{mass_distribution} that 
the lower mass bound of the $Z_2$ boson in the $\psi$ model 
at $\delta = 0$ is much weaker than those of the other $Z'$ 
models. 
It has been pointed out that the most stringent 
constraint on the contact term is the APV measurement 
of cesium atom~\cite{chm}. 
Since  all the SM matter fields in the $\psi$ model 
have the same ${\rm U(1)'}$ charge (see Table~1), 
the couplings of contact interactions are Parity conserving, 
which makes constraint from the APV measurement useless. 
We also find in Fig.~\ref{mass_distribution} that the lower mass 
bound of the $Z_2$ boson disappears near the leptophobic $\eta$-model 
($\beta_E = \tan^{-1}(\sqrt{5/3})$ and $\delta = 1/3$)~\cite{eta_model}. 

We summarize the 95\% CL lower bound on $m_{Z_2}$ 
for the $\chi,\psi,\eta$ and $\nu$ models ($\delta = 0$) 
in Table~\ref{mzelenc}. 
For comparison, we also show the lower bound of $m_{Z_2}$  
in the previous study~\cite{cvetic_langacker_review} 
in the same table. 
The bounds on the $Z_\chi$ and $Z_\nu$ masses 
are more severely constrained 
as compared to ref.~\cite{cvetic_langacker_review}.  
Although we used the latest electroweak data, our result 
for the $Z_\psi$ boson mass is somewhat weaker than that of 
ref.~\cite{cvetic_langacker_review}.  
In the analysis of ref.~\cite{cvetic_langacker_review}, 
the $e^+e^- \to \mu^+\mu^-, 
\tau^+\tau^-$ data below the $Z$ pole 
are also used besides the $Z$-pole, $m_W$ and the LENC data. 
As we mentioned before, the lower mass bound of the $Z'$ boson 
is obtained from the LENC data, not from the $Z$-pole data.  
Because the APV measurement which is most stringent constraint 
in the LENC data does not well constrain the $\psi$ model, 
it is expected that the $e^+ e^-$ annihilation data below 
the $Z$-pole play an important role to obtain the bound of 
$Z_\psi$ boson mass. 
\bound_lenc 

Our results in Table~\ref{mzelenc} are also slightly weaker than 
those in ref.~\cite{chm}. 
The results in ref.~\cite{chm} have been obtained 
without including the $Z$-$Z'$ mixing effects and by setting 
$\mt = 175~{\rm GeV}, \mh = 100~{\rm GeV}, \xa = 0$ 
and $T_{\rm new} = 0$. 

\subsection{Lower mass bound of $Z_2$ boson}
We have found that the $Z$-pole, $m_W$ and the LENC data 
constrain ($T_{\rm new}, \xibar$), $T_{\rm new}$ and $\contact$, 
respectively. 
We can see from eq.~(\ref{eq:zeta}) that, for a given $\zeta$, 
constraints on $T_{\rm new}, \xibar$ and $\contact$ can be 
interpreted as the bound on $m_{Z_2}$. 
We show the 95\% CL lower mass bound of the $Z_2$ boson 
for $\mh = 100~{\rm GeV}$ 
in four $Z'$ models as a function of $\zeta$. 
The bound is again found under the condition $m_{Z_2} \geq 0$. 
Results are shown in Fig.~\ref{mass_95cl}.(a) $\sim$  
\ref{mass_95cl}.(d) for the $\chi,\psi,\eta,\nu$ models, respectively. 
The lower bound from the $Z$-pole and $m_W$ data, and 
that from the LENC data are separately plotted in the same figure. 
In order to see the $g_E$-dependence of the $m_{Z_2}$ bound 
explicitly, we show the lower mass bound for the combination 
$m_{Z_2}g_Y/g_E$. 
We can read off from Fig.~\ref{mass_95cl} that the bound on 
$m_{Z_2}g_Y/g_E$ is approximately independent of $g_E$ for 
$g_E/g_Y = 0.5 \sim 2.0$ in each model. 
\zprime_mass 
As we expected from the formulae for 
$T_{\rm new}$ and $\xibar$ in the small 
mixing limit (eq.~(\ref{eq:tnew_xibar})), 
the $Z_2$ mass is unbounded from the $Z$-pole data at 
$\zeta = 0$. 
For models with very small $\zeta$, the lower bound on 
$m_{Z_2}$, therefore, comes from the LENC experiments 
and the direct search experiment at Tevatron. 
For comparison, we plot the 95\% CL lower bound on 
$m_{Z_2}$ obtained from the direct search experiment~\cite{direct_search} 
in Fig.~\ref{mass_95cl}. 
In the direct search experiment, the $Z'$ decays into the exotic 
particles, \eg, the decays into the light right-handed neutrinos 
which are expected for some models, are not taken into account. 
We summarize the 95\% CL lower bound on $m_{Z_2}$ 
for the $\chi,\psi,\eta$ and $\nu$ models ($\delta = 0$) 
obtained from the low-energy data and 
the direct search experiment~\cite{direct_search} 
in Table~\ref{mzelenc}. 

The lower bound of $m_{Z_2}$ is affected by the Higgs boson mass 
through the $T$ parameter. 
As we seen from eqs.~(\ref{eq:const_chi}) $\sim$ (\ref{eq:const_nu}),  
$T_{\rm new}$ tends to be in the physical region ($T_{\rm new} \geq 0$) 
for large $\mh$ $(\xh)$. 
Then, we find that the large Higgs boson mass decreases the lower 
bound of $m_{Z_2}$. 
For $\zeta = 1$, 
the lower $m_{Z_2}$ bound in the $\chi,\psi,\nu$ ($\eta$) models 
for $\mh = 150~{\rm GeV}$ is weaker than that for $\mh = 100~{\rm GeV}$ 
about 7\% (11\%). 
On the other hand, the Higgs boson with $\mh = 80~{\rm GeV}$ 
makes the lower $m_{Z_2}$ bound in all the $Z'$ models 
severe about 5\% as compared to the case for $\mh = 100~{\rm GeV}$.  
Because $T_{\rm new}$ and $\xibar$ are proportional to 
$\zeta^2$ and $\zeta$, respectively (see eq.~(\ref{eq:tnew_xibar})), 
and it is unbounded at $|\zeta| \simeq 0$, 
the lower bound of $m_{Z_2}$ may be independent of $\mh$ 
in the small $|\zeta|$ region. 
The $\mh$-dependence of the lower mass bound obtained 
from the LENC data is safely negligible. 

It should be noted that,  at $\zeta = 0$, 
only the leptophobic $\eta$-model ($\delta = 1/3$) is 
not constrained from both the $Z$-pole and the low-energy 
data. 
The precise analysis and discussion for the lower mass bound 
of the $Z_2$ boson in the leptophobic $\eta$-model can be 
found in ref.~\cite{uch}. 
It is shown in ref.~\cite{uch} that 
the $\zeta$-dependence of the lower mass bound is slightly 
milder than that of the $\eta$-model with $\delta = 0$ 
in Fig.~\ref{mass_95cl}.(c). 

It has been discussed that the presence of $Z_2$ boson 
whose mass is much heavier than the SM $Z$ boson mass,  
say 1 TeV, may lead to a find-tuning problem to stabilize 
the electroweak scale against the ${\rm U(1)'}$ scale~\cite{drees}. 
The $Z_2$ boson with $m_{Z_2} \leq 1~{\rm TeV}$ for $g_E = g_Y$ 
is allowed by the electroweak data only if $\zeta$ satisfies 
the following condition: 
\bea 
\begin{array}{ll}
-0.6 \lsim \zeta \lsim +0.3 & ~~{\rm for~the}~\chi,\psi,\nu~{\rm models}, 
\\
-0.7 \lsim \zeta \lsim +0.6 & ~~{\rm for~the}~\eta~{\rm model}. 
\end{array}
\label{eq:zeta_condition}
\eea
In principle, the parameter $\zeta$ is calculable, together 
with the gauge coupling $g_E$, once the particle spectrum 
of the $E_6$ model is specified. 
In the next section, we calculate the $\zeta$ parameter in 
several $E_6$ $Z'$ models.

%
%
%
%
\section{Light $Z'$ boson in minimal SUSY $E_6$-models}
\clean
It is known that the gauge couplings are not unified 
in the $E_6$ models with three generations of {\bf 27}. 
In order to guarantee the gauge coupling unification, 
a pair of weak-doublets, $H'$ and $\ov{H'}$, 
should be added into  
the particle spectrum at the electroweak scale~\cite{dienes}. 
They could be taken from ${\bf 27} + {\bf \ov{27}}$ 
or the adjoint representation {\bf 78}. 
The ${\rm U(1)'}$ charges of the additional weak doublets 
should have the same magnitude and opposite sign, $a$ and $-a$,  
to cancel the ${\rm U(1)'}$ anomaly. 
In addition, a pair of the complete SU(5) multiplet such as 
${\bf 5 + \ov{5}}$ can be added without spoiling the unification 
of the gauge couplings~\cite{eta_model, dienes}. 

The minimal $E_6$ model which have three generations of {\bf 27} 
and a pair ${\bf 2} + {\bf \ov{2}}$ depends in principle on the 
three cases; $H'$ has the same quantum number as $L$ or $H_d$ of 
{\bf 27}, or $\ov{H_u}$ of ${\bf\ov{27}}$. 
All three cases will be studied below. 
\e6_extra 
The hypercharge and ${\rm U(1)'}$ charge of 
the extra weak doublets for the $\chi,\psi,\eta,\nu$ models 
are listed in Table~\ref{table:extra_higgs}.  
For comparison, we also show those in the model of Babu 
\etal~\cite{eta_model}, where two pairs of 
${\bf 2 + \ov{2}}$ from {\bf 78} and a pair of ${\bf 3 + \ov{3}}$ 
from ${\bf 27 + \ov{27}}$ are introduced 
to achieve the quasi-leptophobity at the weak scale. 

Let us recall the definition of $\zeta$; 
\beq
\zeta = \frac{g_Z}{g_E}\frac{m_{ZZ'}^2}{m_Z^2} - \delta.  
\label{eq:zetadef}
\eeq
In the minimal model, 
the following eight scalar-doublets can develop VEV to 
give the mass terms $m_Z^2$ and $m_{ZZ'}^2$ in eq.~(\ref{eq:l_gauge}): 
three generations of $H_u,  H_d$, 
and an extra pair, $H'$ and $\ov{H'}$. 
Then, $m_Z^2$ and $m_{ZZ'}^2$ are written in terms of 
their VEVs as 
\bsub
\bea
\mmz &=& \frac{1}{2} g_Z^2\biggl[ 
	 \sum_{i=1}^3 \biggl\{ 
	\langle H_u^i \rangle^2 + \langle H_d^i \rangle^2 
	\biggr\} + \langle H' \rangle^2 
	+ \langle \ov{H'} \rangle^2 \biggr], 
\\
m_{ZZ'}^2 &=& \! g_Z g_E \biggl[ 
	\sum_{i=1}^{3} \biggl\{ 
	-Q_E^{H_u} \langle H_u^i \rangle^2 
	+ Q_E^{H_d} \langle H_d^i \rangle^2 
	\biggr \}
	+ Q_E^{H'} \langle H' \rangle^2 
	- Q_E^{\ov{H'}} \langle \ov{H'} \rangle^2 
	\biggr]\!, 
\eea
\esub
where $i$ is the generation index. 
The third component of the weak isospin $I_3$ for the Higgs 
fields are
\beq
I_3(H_d) = I_3(H') = -I_3(H_u) = -I_3(\ov{H'}) = 1/2. 
\eeq
Taking account of the ${\rm U(1)'}$ charges of 
the extra Higgs doublets, $Q_E^{H'} = - Q_E^{\ov{H'}}$, 
we find from eq.~(\ref{eq:zetadef}) 
\bea
\zeta &=& 2 \frac{\disp{ \sum_{i=1}^3 }\biggl\{  
	-Q_E^{H_u} \langle H_u^i \rangle^2 
	+Q_E^{H_d} \langle H_d^i \rangle^2 
	\biggr \}
	+ Q_E^{H'} \biggl( \langle H' \rangle^2 
	+  \langle \ov{H'}\rangle^2  \biggr) }
	{\disp{ \sum_{i=1}^3 }\biggl\{ 
	\langle H_u^i \rangle^2 + \langle H_d^i \rangle^2 
	\biggr\}
	+ \langle H'\rangle^2 + \langle \ov{H'} \rangle^2 } 
	- \delta.  
\label{eq:zeta_final}
\eea
We note here that the observed $\mu$-decay constant 
leads to the following sum rule 
\bea
v_u^2 + v_d^2 + v_{H'}^2 + v_{\ov{H'}}^2 \equiv 
v^2 = \frac{1}{\sqrt{2}G_F} \approx (246~{\rm GeV})^2,  
\eea
where
\bea
\left.
\begin{array}{rr}
	\disp{\sum_{i=1}^3\langle H_u^i \rangle^2 = \frac{v_u^2}{2},}
	&
	\disp{\sum_{i=1}^3\langle H_d^i \rangle^2 = \frac{v_d^2}{2}, }
	\\
	\disp{\langle H' \rangle^2 = \frac{v_{H'}^2}{2}, }
	&
	\disp{\langle \ov{H'} \rangle^2 = \frac{v_{\ov{H'}}^2}{2}}. 
\end{array}
\right.
\eea
By further introducing the notation
\bsub
\bea
\tan \beta &=& \frac{v_u}{v_d}, 
\\
x^2 &=& \frac{v_{H'}^2 + v_{\ov{H'}}^2}{v^2}, 
\eea
\esub
we can express eq.~(\ref{eq:zeta_final}) as 
\bea
\zeta &=& 2 \biggl\{
	-Q_E^{H_u}(1-x^2)\sin^2\beta +Q_E^{H_d}(1-x^2)\cos^2\beta 
	+Q_E^{H'}x^2
	\biggr\} 
	- \delta. 
\eea
Because $H'$ and $\ov{H'}$ are taken from {\bf 27} + 
{$\bf\ov{27}$}, the ${\rm U(1)'}$ charge of 
$H'$, $Q_E^{H'}$, is identified with that of $L$, 
$H_d$ or $\ov{H_u}$. 

Among all the models, only in the $\chi$-model one can have 
smaller number of matter particles. 
In the $\chi$-model, three generations of the matter fields 
{\bf 16} and  a pair of Higgs doublets make the model 
anomaly free. 
In this case, $\zeta$ is found to be independent of $\tan\beta$: 
\bsub
\bea
\zeta &=& 2 \frac{ Q_E^{H_d} - Q_E^{H_u} \tan^2 \beta  }
	{1 + \tan^2 \beta}
	- \delta 
\\
	&=& 2 Q_E^{H_d} - \delta. 
\eea
\esub

\coeff_rge
Let us now examine the kinetic mixing parameter $\delta$ in each model. 
The boundary condition of $\delta$ at the GUT scale is $\delta = 0$. 
The non-zero kinetic mixing term can arise at low-energy 
scale through the following RGEs: 
\bsub
\bea
\frac{d}{dt} \alpha_i &=& \frac{1}{2\pi}b_i \alpha_i^2,  
\label{eq:rgea}
\\
\frac{d}{dt} \alpha_4 &=& \frac{1}{2\pi}
( b_E + 2 b_{1E} \delta + b_1 \delta^2 ) \alpha_4^2 ,
\label{eq:rgeb}\\
\frac{d}{dt} \delta &=& \frac{1}{2\pi}
( b_{1E} + b_1 \delta ) \alpha_1, 
\label{eq:rgec}
\eea
\label{eq:rge}
\esub
where $i=1,2,3$ and $t=\ln \mu$. 
We define $\alpha_1$ and $\alpha_4$ as 
\bea
\alpha_1 \equiv \frac{5}{3}\frac{g_Y^2}{4\pi}, 
~~~~~~~
\alpha_4 \equiv \frac{5}{3}\frac{g_E^2}{4\pi}. 
\eea
The coefficients of the $\beta$-functions for $\alpha_1, 
\alpha_4$ and $\delta$ are: 
\bea
b_1 = \frac{3}{5} {\rm Tr} (Y^2), 
~~~~
b_E = \frac{3}{5} {\rm Tr} (Q_E^2), 
~~~~
b_{1E} = \frac{3}{5} {\rm Tr} (Y Q_E). 
\eea
From eq.~(\ref{eq:rgec}), we can clearly see that the non-zero $\delta$ 
is generated at the weak scale if $b_{1E} \neq 0$ holds. 
In Table~\ref{table:coeff_rge}, we list $b_1, b_E$ and $b_{1E}$ 
in the minimal $\chi,\psi,\eta,\nu$ models and the $\eta_{\rm BKM}$ 
model~\cite{eta_model}. 
As explained above, the $\chi(16)$ model has three generations 
of {\bf 16}, and the $\chi(27)$ model has three generations of 
{\bf 27}. 
We can see from Table~\ref{table:coeff_rge} that the 
magnitudes of the differences $b_1 - b_2$ and $b_2 - b_3$ are 
common among all the models including the minimal supersymmetric 
SM. 
This guarantees the gauge coupling unification 
at $\mu = m_{GUT} \simeq 10^{16}~{\rm GeV}$. 

\ge_value 
It is straightforward to obtain $g_E(m_{Z_1})$ and 
$\delta(m_{Z_1})$ for each model. 
The analytical solutions of eqs.~(\ref{eq:rgea})$\sim$
(\ref{eq:rgec}) are as follows:
\bsub
\bea
\frac{1}{\alpha_i(m_{Z_1})} &=& \frac{1}{\alpha_{GUT}} 
	+ \frac{1}{2\pi}b_i \ln \frac{m_{GUT}}{m_{Z_1}}, 
\\
\delta(m_{Z_1}) &=& -\frac{b_{1E}}{b_1}\biggl( 
	1 - \frac{\alpha_1(m_{Z_1})}{\alpha_{GUT}} \biggr), 
\\
\frac{1}{\alpha_4(m_{Z_1})} &=& \frac{1}{\alpha_{GUT}} 
	+ \biggl\{ \frac{b_E}{b_1} - 
	\biggl(\frac{b_{1E}}{b_1} \biggr)^2 
	\biggr \} 
	  \biggl\{ \frac{1}{\alpha_1(m_{Z_1})} - \frac{1}{\alpha_{GUT}}
	\biggr \} 
\nonumber \\
	&& 
	- \biggl( \frac{b_{1E}}{b_1} \biggr)^2 
	\frac{ \alpha_1(m_{Z_1}) - \alpha_{GUT}}{\alpha_{GUT}^2}, 
\eea
\esub
where $\alpha_{GUT}$ denotes the unified gauge coupling at 
$\mu = m_{GUT}$. 
In our calculation, 
$\alpha_3(m_{Z_1}) = 0.118$ and 
$\alpha(m_{Z_1}) = e^2(m_{Z_1})/4\pi = 1/128$ 
are used as example. 
These numbers give $g_Y(m_{Z_1}) = 0.357$. 
We summarize the predictions for $g_E$ 
and $\delta$ at $\mu = m_{Z_1}$ in the 
minimal $E_6$ models and the $\eta_{\rm BKM}$ model 
in Table~\ref{table:ge_delta}. 
In all the minimal models, the ratio $g_E/g_Y$ is approximately 
unity and $|\delta|$ is smaller than about 0.07. 
On the other hand, the $\eta_{\rm BKM}$ model 
predicts 
\tnb_zeta
$g_E/g_Y \sim 0.86$ and $\delta \sim 0.29$, which is close 
to the leptophobic-$\eta$ model at $\delta = 1/3$. 
In Figs.~\ref{chisq_distribution} and \ref{mass_distribution}, 
we show the predictions of all the models by ``$\times$'' symbol.

Next, we estimate the parameter $\zeta$ for several 
sets of $\tan\beta$ and $x$. 
In Table~\ref{table:zetasummary}, we show the predictions for 
$\zeta$ in the minimal $\chi,\psi,\eta,\nu$ models and the 
$\eta_{\rm BKM}$ model. 
The results are shown for $\tan\beta = 2$ and $30$, and 
$x^2 = 0$ and $0.5$. 
We find from the table that the parameter $\zeta$ is 
in the range $|\zeta| \lsim 1.35$ for all the models 
except for the $\eta_{\rm BKM}$ model, where the 
predicted $\zeta$ lies between $-2.0$ and $-1.2$.  
It is shown in Fig.~\ref{mass_95cl}
that $m_{Z_2}g_Y/g_E$ is approximately 
independent of $g_E/g_Y$.  
Actually, we find in Table~\ref{table:ge_delta} and 
Table~\ref{table:zetasummary} that 
the predicted $|\delta|$ is smaller than about 0.1 
and $g_E/g_Y$ is quite close to unity in all the minimal models.  
We can, therefore, read off from Fig.~\ref{mass_95cl} 
the lower bound of $m_{Z_2}$ in the minimal models at $g_E = g_Y$. 
In Table~\ref{table:mass95_zeta}, we summarize the 95\% CL lower 
$m_{Z_2}$ bound for the minimal $\chi,\psi,\eta,\nu$ models and 
the $\eta_{\rm BKM}$ model which correspond to the predicted $\zeta$ 
in Table~\ref{table:zetasummary}. 
\masszeta
Most of the lower mass bounds in Table~\ref{table:mass95_zeta} 
exceed 1 TeV. 
The $Z_2$ boson with $m_{Z_2} \sim O(1~{\rm TeV})$ should be 
explored at the future collider such as LHC. 
The discovery limit of the $Z'$ boson in the $E_6$ models at LHC 
is expected as~\cite{cvetic_bound}
\bea
\begin{array}{cccc}\hline 
\chi & \psi & \eta & \nu \\ \hline 
3040 & 2910 & 2980 & *** \\ \hline 
\end{array}
\eea
All the lower bounds of $m_{Z_2}$ listed in 
Table~\ref{table:mass95_zeta} are smaller than 2 TeV 
and they are, therefore, in the detectable range of LHC. 
But, it should be noticed that most of them 
($1~{\rm TeV} \lsim m_{Z_2}$) may require the fine-tuning to stabilize 
the electroweak scale against the ${\rm U(1)'}$ scale~\cite{drees}. 

The lower bound of the $Z_2$ boson mass in the $\eta_{\rm BKM}$ 
model for the predicted $\zeta$ can be read off from Fig.~2 in 
ref.~\cite{uch}. 
Because somewhat large $\zeta$ is predicted in the $\eta_{\rm BKM}$ 
model, $1 \lsim |\zeta|$, the lower mass bound is also large 
as compared to the minimal models. 

%
%
%
%
\section{Summary}
We have studied constraints on $Z'$ bosons in the SUSY 
$E_6$ models. 
Four $Z'$ models --- the $\chi,\psi,\eta$ and $\nu$ models 
are studied in detail. 
The presence of the $Z'$ boson affects the electroweak processes 
through the effective $Z$-$Z'$ mass mixing angle $\xibar$, 
a tree level contribution $T_{\rm new}$ which is a positive 
definite quantity, and the contact term $\contact$. 
The $Z$-pole, $m_W$ and LENC data constrain ($T_{\rm new}, 
\xibar$), $T_{\rm new}$ and $\contact$, respectively. 
The convenient parametrization of the electroweak 
observables in the SM and the $Z'$ models are presented. 
From the updated electroweak data, we find that the $Z'$ models 
never give the significant improvement of the $\chi^2$-fit 
even if the kinetic mixing is taken into accounted. 
The 95\% CL lower mass bound of the heavier mass eigenstate $Z_2$ 
is given as a function of the effective $Z$-$Z'$ mixing 
parameter $\zeta$. 
The approximate scaling low is found for the $g_E/g_Y$-dependence 
of the lower limit of $m_{Z_2}$. 
By assuming the minimal particle content of the $E_6$ model,  
we have found the theoretical predictions for $\zeta$. 
We have shown that the $E_6$ models 
with minimal particle content which is consistent with 
the gauge coupling unification predict the non-zero kinetic mixing 
term $\delta$ and the effective mixing parameter $\zeta$ of 
order one. 
The present electroweak experiments lead to the lower 
mass bound of order 1 TeV or larger for those models. 

%
%
%
%
\section*{Acknowledgment}
This work is supported in part by Grant-in-Aid for Scientific 
Research from the Ministry of Education, Science and Culture of Japan.

%
%
%
%
%
%

\end{document}